\documentclass[%twocolumn,%showpacs,preprintnumbers,
amsmath,%amssymb,aps,
prd,nofootinbib,floatfix,11pt,%preprint,
]{revtex4}

\usepackage{placeins}
\usepackage{afterpage}

\usepackage{graphicx}% Include figure filesf
\usepackage{setspace}
\usepackage{bm}% bold math
\newcommand\beq{\begin{eqnarray}}
\newcommand\eeq{\end{eqnarray}}

\def\lsim{\mathrel{\rlap{\lower4pt\hbox{$\sim$}}
    \raise1pt\hbox{$<$}}}                % less than or approx. symbol
\def\gsim{\mathrel{\rlap{\lower4pt\hbox{$\sim$}}
    \raise1pt\hbox{$>$}}}            

\allowdisplaybreaks
\begin{document}

\renewcommand{\theequation}{\arabic{section}.\arabic{equation}}
\renewcommand{\thefigure}{\arabic{section}.\arabic{figure}}
\renewcommand{\thetable}{\arabic{section}.\arabic{table}}

\title{\Large \baselineskip=20pt 
Implications of purity constraints on light higgsinos
}
\author{Stephen P.~Martin}
\affiliation{\mbox{\it Department of Physics, Northern Illinois University, DeKalb IL 60115}}

\begin{abstract}\normalsize \baselineskip=15.5pt 
The lightest supersymmetric particles could be higgsinos that have a small mixing with gauginos. If the lightest higgsino-like state makes up some or all of the dark matter with a thermal freezeout density, then its mass must be between about 100 and 1150 GeV, and dark matter searches put bounds on the amount of gaugino contamination that it can have. Motivated by the generally good agreement of flavor- and CP-violating observables with Standard Model predictions, I consider models in which the scalar particles of minimal supersymmetry are heavy enough to be essentially decoupled, except for the 125 GeV Higgs boson. I survey the resulting purity constraints as lower bounds on the gaugino masses and upper bounds on the higgsino mass splittings. I also discuss the mild excesses in recent soft lepton searches for charginos and neutralinos at the LHC, and show that they can be accommodated in these models if $\tan\beta$ is small and $\mu$ is negative.
\end{abstract}

\maketitle

\tableofcontents

\baselineskip=15.6pt

\newpage

%%%%%%%%%%%%%%%%%%%%%%%%%%%%%%%%%%%%%%%%%%%%%%%%%%%%%%%%%%%%%%%
\section{Introduction\label{sec:intro}}
\setcounter{equation}{0}
\setcounter{figure}{0}
\setcounter{table}{0} 
\setcounter{footnote}{1}

Softly broken supersymmetry (for reviews in conventions and notations consistent with that of the present paper, see refs.~\cite{Martin:1997ns,Dreiner:2023yus}) provides a solution to the large hierarchy problem associated with the small ratio of the electroweak mass scale to the Planck scale and other very high mass scales. The radiative correction sensitivity of the electroweak scale to arbitrarily large mass scales is reduced to a sensitivity to the scale of soft supersymmetry breaking, which is presumed to be many orders of magnitude smaller due to a dynamical origin, for which there is no shortage of proposals.

On the other hand, supersymmetry breaking introduces new possible sources of flavor violation and CP violation. Furthermore, the continuing explorations of the Large Hadron Collider (LHC) are putting lower bounds on superpartner masses, especially in the cases of the strongly interacting superpartners, gluinos and squarks, which are produced with relatively large cross-sections in proton-proton collisions.  In recent years, search results are mostly presented in terms of simplified models, and it should be recognized that the simplified models of supersymmetry are not actually supersymmetry. The quoted bounds are therefore often significantly stronger than what would follow from realistic supersymmetric models. Still, the indirect bounds from flavor and CP violation together with the direct bounds from collider searches motivate a scenario \cite{Wells:2003tf,Arkani-Hamed:2004zhs,Wells:2004di,Arvanitaki:2012ps}
with a little hierarchy, in which the squarks and sleptons would have multi-TeV or even PeV masses, beyond the reach of the LHC. The proven existence of a Higgs scalar boson with mass near $M_h = 125.1$ GeV gives credence to this possibility, since the tree-level prediction for the mass is less than $M_Z$, and so it is both necessary and sufficient to have logarithmic corrections to $M_h$ that can easily have the right magnitude if top squarks have multi-TeV or larger masses.

This raises the question of what superpartners could most plausibly still appear at the LHC, and what type of neutral lightest supersymmetric particle (LSP) could be detected in dark matter searches. The accumulation of data at the LHC has arguably reached the point that gluino and squark search improvements are more incremental, and the largest improvements in reach will come in the cases of charginos and neutralinos, which are mixtures of the superpartners of the Higgs bosons (higgsinos) and the gauge bosons of the Standard Model (gauginos). Many studies have pointed out the intriguing features of light higgsino-like particles, for which the present LHC bounds are rather weak, and dark matter bounds have not reached the exclusion level. The electroweak scale in supersymmetry is closely tied to the magnitude of the higgsino mass parameter $\mu$. This makes it interesting to consider the case that among the new mass parameters required in the Minimal Supersymmetric Standard Model (MSSM), $\mu$ is closest to the electroweak scale, with other superpartner masses (perhaps much) larger. Furthermore, a nearly pure higgsino LSP would make a good candidate for the dark matter that seems to be required by cosmology and astrophysics observations (for some recent reviews, see refs.~\cite{Roszkowski:2017nbc,Feng:2022rxt,Safdi:2022xkm}). 

If it makes up a significant fraction of the dark matter, an extremely pure  vectorlike spin-1/2 fermion in a doublet of $SU(2)_L$ with weak hypercharge $\pm 1/2$ with a mass splitting between its two neutral mass eigenstates of less than about 200 keV would be ruled out by direct searches from inelastic scattering of the LSP into the higher neutral state, mediated by the $Z$ boson. Fortunately, this kind of simplified ``higgsino" model is not at all what appears in realistic supersymmetric models, which essentially always have a much larger mass splitting due to the mixing with the gauginos. This impurity of higgsino-like neutralinos in supersymmetry is both necessary and guaranteed,
assuming the gaugino masses are not enormous, since the off-diagonal terms in the neutralino mass matrix in the gauge-eigenstate basis have magnitudes that are bounded from below by electroweak symmetry breaking.

In this paper, I consider a class of supersymmetric models, defined by parameters at the 10 TeV scale (rather than the GUT or Planck scale), in which the lightest superpartners are two higgsino-like neutralinos and a chargino, denoted $\tilde N_1$, $\tilde N_2$, and $\tilde C_1$. These particles have mixing with the bino and wino, 
yielding further neutralino and chargino states $\tilde N_3$, $\tilde N_4$, and $\tilde C_2$, which are assumed here to be at least somewhat heavier. Inspired by the arguments of \cite{Wells:2003tf,Arkani-Hamed:2004zhs,Wells:2004di,Arvanitaki:2012ps}, and by the lack of discovery from the LHC,
and by the Higgs boson mass of 125.1 GeV, the squarks, sleptons, gluino, and the other Higgs bosons $A, H, H^\pm$ will be assumed to be practically decoupled, with a common default mass of 10 TeV. This provides a simple framework with only a few parameters having a significant impact on immediately relevant physics considerations, namely the higgsino mass parameter $\mu$, the bino and wino masses $M_1$ and $M_2$, which are assumed to be real and larger in magnitude than $|\mu|$,
and $\tan\beta$, the ratio of the Higgs expectation values. Besides simplicity, this framework has the virtues of avoiding both direct detection of superpartners at the LHC and indirect constraints from flavor and CP-violating observables.

\begin{figure}[!t]
\begin{minipage}[]{0.5\linewidth}
\includegraphics[width=\linewidth]{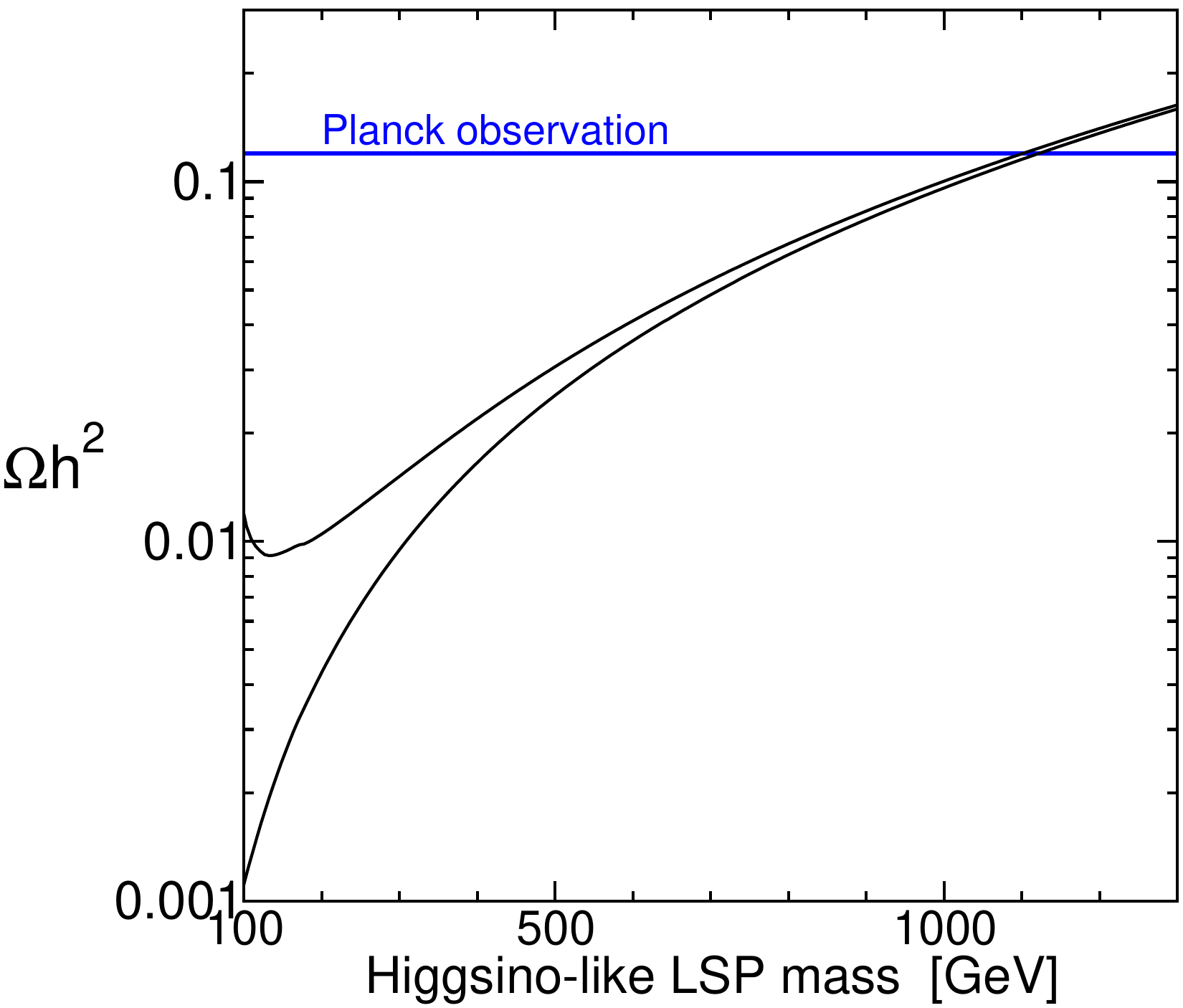}
\end{minipage}
\hspace{0.05\linewidth}
\begin{minipage}[]{0.42\linewidth}
\caption{\label{fig:Omegah2}
The thermal freezeout prediction for $\Omega h^2$ for higgsino-like LSP dark matter, 
as a function of the mass. For the illustration, the squark, slepton, and $H,A,H^\pm$
Higgs boson masses are set to 10 TeV. The upper black curve assumes $\mu : M_1 : M_2 = 1 : 1.5 : 3$ and $\tan\beta=2$,
while the lower black curve assumes $M_1 = M_2 = 10$ TeV and $\tan\beta=10$.
In both cases, agreement with the Planck experiment result \cite{Planck:2018vyg}
of $\Omega h^2 = 0.12$ is obtained if the LSP mass $M_{\tilde N_1} \approx 1.1$ TeV,
but significant differences arise for smaller masses.}
\end{minipage}
\end{figure}

It is well-known that the present abundance of a higgsino-like LSP, assuming thermal freezeout in the standard cosmology, increases with the mass, and will agree with the best fit of $\Omega_{\rm DM} h^2 = 0.12$ from the results of the Planck experiment \cite{Planck:2018vyg} if $M_{\tilde N_1}$ is about 1.1 TeV. This is illustrated in 
Figure \ref{fig:Omegah2}, which shows the thermal freezeout prediction for $\Omega h^2$
as a function of $M_{\tilde N_1}$, as computed by the public code {\tt micrOMEGAs v6.0} 
\cite{Belanger:2001fz,Belanger:2004yn,Belanger:2020gnr,Alguero:2023zol}, for two cases
with decoupled and not-so-decoupled gauginos. This shows that $\Omega h^2$ has more sensitivity to the gaugino mixing when it is small (for lighter higgsinos, with $|\mu|$ closer to $m_Z$).

%%%%%%%%%%%%%%%%%%%%%%%%%%%%%%%%%%%%%%%%%%%%%%%%%%%%%%%%%%%%%%%%%%%%%%%%%%%%%%%%%
%%%%%%%%%%%%%%%%%%%%%%%%%%%%%%%%%%%%%%%%%%%%%%%%%%%%%%%%%%%%%%%%%%%%%%%%%%%%%%%%%

At least four obvious possibilities present themselves for a higgsino-like LSP. First, the dark matter could indeed be a higgsino-like LSP with mass near 1.1 TeV. Second, the mass could be less than 1.1 TeV, and the remaining dark matter could be something else, for example axions. In this case, the rates for dark matter direct detection experiments, for a given LSP-nucleon cross-section will be reduced by a factor of
\beq
\xi \equiv 
\frac{\Omega_{\rm LSP} h^2}{0.12} \leq 1,
\eeq
while rates for most indirect detection searches (with the exception of searches
for annihilations to neutrinos from equilibrium densities accumulated in the Sun) will be reduced by $\xi^2$.
Third, the mass could again be less than 1.1 TeV, but with other non-thermal-freezeout sources for the higgsino dark matter. Fourth, the higgsino might be the lightest among the subset of supersymmetric particles within the MSSM, but it decays to something else, either to Standard Model states through $R$-parity violating couplings, or to non-MSSM $R$-parity-odd states such as a singlino (perhaps an axino). In this paper, I concentrate on the first two of these possibilities.
The proposal that a higgsino LSP and an axion combine to give the dark matter density inferred by cosmology has been studied in detail in
refs.~\cite{Bae:2013bva,Bae:2014yta,Bae:2017hlp,Baer:2019uom}, with an emphasis on fine-tuning considerations that will play no role in the present paper.
Other important contributions to our understanding of the higgsino LSP scenario, including the difficulties in observing them in collider experiments and their possible role as dark matter, can be found in
refs.~\cite{Drees:1996pk,Thomas:1998wy,Giudice:2004tc,Profumo:2004at,Hisano:2004ds,Baer:2011ec,Baer:2012cf,Baer:2014cua,Nagata:2014wma,Evans:2014pxa,Bae:2015jea,Fukuda:2017jmk,Kowalska:2018toh,Baer:2018rhs,Fukuda:2019kbp,Baer:2020sgm,Rinchiuso:2020skh,Delgado:2020url,Co:2021ion,Baer:2021srt,Carpenter:2021jbd,Evans:2022gom,Baer:2022qrw,Dessert:2022evk,Carpenter:2023agq,Bisal:2023fgb,Ibe:2023dcu}.

Dark matter searches of several types impose constraints on the higgsino-like LSP parameter space. These constraints are non-trivial even if the mass is less than 1.1 TeV and thermal freezeout governs the density, so that $\xi$ is less than 1.
Annihilations of dark matter have been searched for using gamma rays from the center of our galaxy with the HESS ground-based Cherenkov telescope array \cite{HESS:2016mib,HESS:2018cbt,HESS:2022ygk},
and from annihilations in dwarf galaxies
with the Fermi-LAT space-based gamma-ray telescope  
\cite{Fermi-LAT:2015ycq,Fermi-LAT:2015att,MAGIC:2016xys}. At present, gamma-ray searches do not exclude the 1.1 TeV thermal higgsino case \cite{Dessert:2022evk,Ellis:2022emx}, 
and are subject to significant uncertainties regarding the dark matter density profile, and become weaker proportionally to $\xi^2$. In contrast, searches by the IceCube experiment for
neutrinos from annihilations in the Sun \cite{IceCube:2016dgk} depend on the equilibrium rate of dark matter accumulations and are proportional to $\xi$, and give constraints on the spin-dependent LSP-proton cross-section. However, at present
the preceding constraints on higgsino-like dark matter with a thermal freezeout relic abundance are less stringent than those coming from the
LUX-ZEPLIN \cite{LZ:2022lsv} (LZ 2022) direct detection searches, which limit both the
spin-independent LSP-xenon nuclear cross-section, and (in certain parts of parameter space that may be quite important, as we will see) the LSP-neutron spin-dependent cross-section. These constraints scale proportional to $\xi$, but are non-trivial even for higgsino-like LSPs with masses close to 100 GeV, so that $\xi$ is as low as of order $0.01$. In general, such direct detection constraints have the effect of requiring that the higgsino LSP is sufficiently pure (bounding the gaugino contents of the LSP mass eigenstate) in different ways, and to different extents, depending on the mass and other parameters.

In this paper, I will survey the requirements on light higgsino purity that follow from the LZ 2022 bounds, using the model framework with decoupled scalars outlined above. For fixed values of $\mu$ and $\tan\beta$, I obtain lower bounds on the gaugino masses, with consequences for the collider search prospects for the bino-like and wino-like neutralinos and charginos. These constraints also entail upper bounds on the mass splittings between the mostly higgsino states $\tilde N_1$ (the LSP) and $\tilde N_2$ and $\tilde C_1$. This has implications for past and future LHC searches, and I will comment in particular on the slight excesses in soft lepton searches 
observed by the ATLAS and CMS collaborations
\cite{ATLAS:2021moa,CMS:2023qhl,CMS:2021edw}. 

\baselineskip=14.9pt

\section{Higgsino mixing and couplings to Higgs and $Z$ bosons\label{sec:expansion}}
\setcounter{equation}{0}
\setcounter{figure}{0}
\setcounter{table}{0} 
\setcounter{footnote}{1}

In this section, I review the tree-level mixing and mass splitting of neutralinos and charginos in the MSSM. Although these tree-level results are not completely adequate and are not used for the detailed studies in the remaining sections, they can be used to help understand certain qualitative features, notably the dependences on $\tan\beta$ and the sign of $\mu$.

In the gauge eigenstate basis $(\tilde B, \tilde W^0, \tilde H_d^0, \tilde H_u^0)$,
the tree-level neutralino mass matrix can be divided into parts that do not and do
rely on electroweak symmetry breaking,
\beq
M_{\tilde N} &=& 
\begin{pmatrix} 
M_1 & 0 & 0 & 0 \\
0 & M_2 & 0 & 0 \\
0 & 0 & 0 & -\mu \\
0 & 0 & -\mu & 0
\end{pmatrix}
+ m_Z
\begin{pmatrix}
0 & 0 & -s_W c_\beta & s_W s_\beta \\
0 & 0 & c_W c_\beta & -c_W s_\beta \\
-s_W c_\beta & c_W c_\beta & 0 & 0 \\
s_W s_\beta & -c_W s_\beta & 0 & 0 
\end{pmatrix}
,
\eeq
where $s_W$ and $c_W$ are the sine and cosine of the weak mixing angle, and $s_\beta$ and $c_\beta$ are the sine and cosine of the angle $\beta$ defined by $\tan\beta = \langle H_u^0 \rangle/\langle H_d^0 \rangle$, the ratio of Higgs expectation values. The neutralino mixing matrix with elements $N_{jk}$ is conventionally defined so that
\beq
N^* M_{\tilde N} N^{-1} = {\rm diag}(M_{\tilde N_1}, M_{\tilde N_2}, M_{\tilde N_3}, M_{\tilde N_4}),
\eeq
and in our case $\tilde N_1$ is the LSP and $\tilde N_2$ is the other higgsino-like fermion, and the mass eigenvalues $M_{\tilde N_j}$ are always real and positive (no matter what the phases of $M_1$, $M_2$, and $\mu$ are). For simplicity, suppose
that $M_1$, $M_2$, are real and positive, and that $\mu$ is also real with 
\beq
\sigma \,\equiv\, \mbox{sign}(\mu) \,=\, \pm 1.
\eeq
Doing perturbation theory in $m_Z$, one finds that the tree-level mass eigenvalues are
(neglecting contributions of order $m_Z^4$)
\beq
M_{\tilde N_1} &=& |\mu| - \frac{m_Z^2}{2} (1 + \sigma s_{2\beta}) \left (
\frac{c_W^2}{M_2 - |\mu|} + \frac{s_W^2}{M_1 - |\mu|}
\right ) + \cdots,
\\
M_{\tilde N_2} &=& |\mu| + \frac{m_Z^2}{2} (1 - \sigma s_{2\beta}) \left (
\frac{c_W^2}{M_2 + |\mu|} + \frac{s_W^2}{M_1 + |\mu|}
\right ) + \cdots,
\\
M_{\tilde N_3} &=& M_1 + s_W^2 m_Z^2 \left (\frac{M_1 + \mu s_{2\beta}}{M_1^2 - \mu^2}\right ) + \cdots,
\\
M_{\tilde N_4} &=& M_2 + c_W^2 m_Z^2 \left (\frac{M_2 + \mu s_{2\beta}}{M_2^2 - \mu^2}\right ) + \cdots,
\eeq
where $s_{2\beta} = \sin(2\beta)$,
so that $\tilde N_1$ is the LSP for $M_1, M_2 > |\mu|$. 
The corresponding neutralino
mixing matrix can also be straightforwardly evaluated in perturbation theory to order $m_Z^2$. Of particular interest are the resulting LSP couplings to the lightest Higgs boson $h$ and to the $Z$ boson, which respectively give the most important contributions
to the spin-independent and spin-dependent LSP-nucleon cross-sections, given the assumption of decoupled squarks, sleptons, and $A, H, H^\pm$ bosons.
They have the form
\beq
{\cal L} &=& 
-\sqrt{g^2 + g^{\prime 2}} \left ( \frac{1}{2} y_{h}^{\phantom{Z}} h \tilde N_1 \tilde N_1 + {\rm c.c.}  + g_{Z}^{\phantom{Z}} 
Z_\mu \tilde N_1^\dagger \overline\sigma^\mu \tilde N_1 \right )
,
\eeq 
where
\beq
y_h^{\phantom{Z}} &=& \left (c_W N_{12}^* - s_W N_{11}^* \right )
(-s_\alpha N_{13}^* - c_\alpha N_{14}^*), 
\\
g_Z^{\phantom{Z}} &=& (|N_{14}|^2 - |N_{13}|^2)/2,
\eeq
in which $\alpha$ is the mixing angle in the neutral Higgs scalar boson sector.
Taking the decoupling limit in which $\alpha = \beta - \pi/2$, one finds the following results at leading order in an expansion in $m_Z$,
\beq
y_h^{\phantom{Z}} &=& -\frac{m_Z}{2} (1 + \sigma s_{2\beta})
\left ( \frac{c_W^2}{M_2 - |\mu|} + \frac{s_W^2}{M_1 - |\mu|} \right ),
\label{eq:yh}
\\
g_Z^{\phantom{Z}} &=& \frac{m_Z^2}{4|\mu|} c_{2\beta} \left ( \frac{c_W^2}{M_2 - |\mu|} + \frac{s_W^2}{M_1 - |\mu|} \right ). 
\label{eq:gZtree}
\eeq
These both vanish in the limit that $m_Z$ is small compared to $M_2 - |\mu|$ and $M_1 - |\mu|$, corresponding to a pure higgsino LSP. Two more comments concerning the LSP coupling to the Higgs boson are in order. First, for fixed values of $\tan\beta$ and sign($\mu$), the coupling has a very simple relation to the LSP mass shift, namely
\beq
y_h^{\phantom{Z}} &=& (M_{\tilde N_1} - |\mu|)/{m_Z}
.
\eeq
Second, for fixed $M_1$, $M_2$, and $|\mu|$, the magnitude of $y_h$ is a monotonically
increasing function of $\sigma/\tan\beta$ in the physical range of $\tan\beta > 1$. This means that when all other Higgs bosons and squarks are decoupled, the spin-independent LSP-nucleon coupling, proportional to $y_h^2$, is minimized for small $\tan\beta$ and negative $\mu$, maximized for small $\tan\beta$ and positive $\mu$, and intermediate for larger $\tan\beta$. Indeed, in the formal limit of $\tan\beta \rightarrow 1$ for negative $\mu$, both $y_h$ and $g_Z$ vanish, corresponding \cite{Cheung:2012qy} to blind spots\footnote{Other neutralino direct detection blind spots have been noted in e.g.~refs.~\cite{Ellis:2000ds,Baer:2003jb}.} in both the  spin-independent and spin-dependent direct detection cross-sections.
However, $\tan\beta$ too close to 1 is problematic because it would require a top-quark Yukawa coupling larger than its MSSM infrared fixed-point value, leading to non-perturbative behavior at high energies unless some other new physics intervenes. For somewhat larger $\tan\beta$ and $\mu<0$, the suppression of the factor 
$1 + \sigma s_{2\beta}$ for $y_h$ is significantly more robust than that of the factor $c_{2\beta}$ for $g_Z$. Another potential issue with small $\tan\beta$ is that even for rather large
top-squark masses and significant top-squark mixing, it may be difficult to achieve $M_h = 125.1$ GeV in the MSSM. This can be circumvented, if necessary, by introducing vectorlike quark supermultiplets with large Yukawa couplings \cite{Moroi:1992zk,Moroi:1991mg,Babu:2004xg,Babu:2008ge,Martin:2009bg,Graham:2009gy}, which can substantially raise the lightest Higgs boson mass without affecting other considerations in this paper.

The spin-dependent LSP-nucleon cross-sections in the decoupling limit includes a contribution proportional to the coupling $g_Z^2$. Equation~(\ref{eq:gZtree}) shows that the magnitude of $g_Z$ is also proportional to $M_{\tilde N_1} - |\mu|$, but is suppressed by an additional factor of $m_Z/|\mu|$ compared to $y_h$, so that it has relatively enhanced importance for lighter higgsinos compared to heavier higgsinos. It also has a different $\tan\beta$ dependent prefactor, which makes it potentially relatively more important than $y_h$ for direct detection for smaller $\tan\beta$ when $\mu$ is negative. This is indeed the case as we will see.

The tree-level chargino mass matrix is given in the gauge-eigenstate basis by
\beq
M_{\tilde C} = \begin{pmatrix}
0 & X^T \\
X & 0\end{pmatrix},
\qquad\quad
X = \begin{pmatrix}
M_2 & \sqrt{2} s_\beta c_W m_Z \\
\sqrt{2} c_\beta c_W m_Z & \mu
\end{pmatrix},
\eeq
leading to masses
\beq
M_{\tilde C_1} &=& |\mu| - m_Z^2 c_W^2 \left ( \frac{|\mu| + M_2 \sigma s_{2\beta}}{M_2^2 - \mu^2}\right ) + \cdots
,
\\
M_{\tilde C_2} &=& M_2 + m_Z^2 c_W^2  \left (\frac{M_2 + \mu s_{2\beta}}{M_2^2 - \mu^2}\right ) 
+ \cdots.
\eeq
It follows that the higgsino-like fermion mass splittings at tree-level are
\beq
&&\!\!\!\!\!\!
\Delta M_0 \,\equiv\, 
M_{\tilde N_2} - M_{\tilde N_1} 
\nonumber
\\ 
&&= 
\frac{m_Z^2}{2} \left \{ [1 + \sigma s_{2\beta}] \left (
\frac{c_W^2}{M_2 - |\mu|} + \frac{s_W^2}{M_1 - |\mu|}
\right )
+ 
[1 - \sigma s_{2\beta}] \left (
\frac{c_W^2}{M_2 + |\mu|} + \frac{s_W^2}{M_1 + |\mu|}
\right )
\right \} + \ldots
\phantom{xxx}
\eeq
for the neutralinos, and
\beq
&&\!\!\!\!\!
\Delta M_+ \,\equiv\, 
M_{\tilde C_1} - M_{\tilde N_1} 
\nonumber
\\ 
&&= 
\frac{m_Z^2}{2} \left \{ [1 + \sigma s_{2\beta}] \left (
\frac{c_W^2}{M_2 - |\mu|} + \frac{s_W^2}{M_1 - |\mu|}
\right )
- 2 c_W^2 \left ( \frac{|\mu| + M_2 \sigma s_{2\beta}}{M_2^2 - \mu^2}\right )
\right \} + \ldots
\label{eq:deltaMp}
\eeq
for the chargino-LSP mass difference, where the ellipses represent higher orders in the expansion in $m_Z$. If one now takes the further limit $M_1, M_2 \gg |\mu|$,
then we have
%, for $m_Z \ll |\mu| \ll M_1, M_2$,
\beq
\Delta M_0 &=& m_Z^2 \left (\frac{c_W^2}{M_2} + \frac{s_W^2}{M_1} \right ) + \ldots
\\
\Delta M_+ &=& \frac{m_Z^2}{2} \left (
[1 - \sigma s_{2\beta}] \frac{c_W^2}{M_2} + 
[1 + \sigma s_{2\beta}] \frac{s_W^2}{M_1} \right ) 
+ \ldots,
\eeq
and then in the further limit of large $\tan\beta$, or when $M_1/M_2 = s_W^2/c_W^2$, one finds $\Delta M_0 = 2 \Delta M_+$,
the relation that has been used in LHC experimental papers \cite{ATLAS:2021moa,CMS:2021edw} 
to define simplified models of higgsinos used for quoting search bounds. However, it is important that this relation is not always satisfied even approximately, since the relevant hierarchies used to derive it need not hold, even when the higgsinos are rather pure.
 
Furthermore, the preceding discussion involves the tree-level neutralino mass matrix, but radiative corrections can be important, especially when the mass splittings are small and/or when there is a significant hierarchy between the gaugino masses and $|\mu|$. In the extreme case of a pure higgsino with the gauginos completely decoupled so that the $m_Z$ corrections can be neglected, the higgsinos would all be degenerate at tree-level.
However, it was shown in ref.~\cite{Thomas:1998wy} that the charginos get a positive
radiative mass splitting from Standard Model electroweak gauge interactions, so 
\beq
\Delta M_0 = 0,\qquad
\Delta M_+ = F(|\mu|/m_Z)\> \mbox{355 MeV} , 
\label{eq:thomaswells}
\eeq
where the loop integral function
\beq
F(x) = \frac{x}{\pi} \int_0^1 dt\>\, (2 - t) \ln [1 + t/x^2 (1-t)^2],
\eeq
monotonically increases from about $F(1.10) =  0.724$ for light higgsinos with mass 100 GeV to an asymptotic value $F(\infty) = 1$ for heavy higgsinos. Equation (\ref{eq:thomaswells}) is used, as a simplified model, in the presentation of experimental search results for quasi-stable charged higgsinos manifesting as disappearing tracks \cite{ATLAS:2022rme,CMS:2023mny},
but it should again be recognized that it will not be generically realized in actual supersymmetric models unless the gaugino masses are of order 10 TeV or much more, depending on the other parameters. For smaller gaugino masses, the tree-level contributions to the mass splittings discussed above are comparable or much larger
than this radiative correction.

Another way of estimating the radiative corrections to the higgsino mass splittings
was put forward in ref.~\cite{Nagata:2014wma}, which integrates out the gauginos
to define an effective field theory below the gaugino mass scale set by $M_1, M_2$.
This approach works well
as long as $M_1, M_2 \gg |\mu|$, and has the advantage of resumming  
logarithmic correction to the masses using the renormalization group in the effective field theory. However, in the present paper, I will be interested in cases where $|\mu|$
is not so small compared to at least one of the gaugino masses. Therefore, below I choose to use the full 1-loop corrections to the neutralino and chargino masses, which only includes the leading logarithms but does not neglect subleading
terms in $|\mu|/M_1$ and $|\mu|/M_2$ including the tree-level ones. The analytic form of these corrections can be found in refs.~\cite{Pierce:1994ew,Pierce:1996zz,Eberl:2001eu,Martin:2005ch}
 in various schemes, and I use the software implementation in the public code 
{\tt SOFTSUSY v4.1.12} \cite{Allanach:2001kg}, which interfaces to {\tt micrOMEGAs}. I find that
the results of {\tt SOFTSUSY} are in reasonably good agreement with the analytic results
found in \cite{Martin:2005ch}, which in turn approach the results of
ref.~\cite{Thomas:1998wy} in the extreme limit $|\mu| \ll M_1, M_2$ when all scalars except $M_h=125.1$ GeV are decoupled.

\section{Gaugino mass bounds for higgsino dark matter with decoupled scalars\label{sec:inobounds}}
\setcounter{equation}{0}
\setcounter{figure}{0}
\setcounter{table}{0} 
\setcounter{footnote}{1}

The interaction of the LSP with ordinary matter, like the xenon nuclei used in the LZ experiment, relies on the predominantly higgsino-like state mixing with gauginos. In the limit that the gaugino mass parameters become large, both the spin-independent and spin-dependent cross-sections become small.
This implies that the observed direct detection limits give lower bounds on the gaugino masses $M_1$ and $M_2$, in order to avoid too much mixing. In most of the parameter space,
the spin-independent LSP nucleon cross-section $\sigma_{\rm SI}$ (corresponding to proton and neutron components weighted by their numbers in the xenon nucleus) give the strongest bound from LZ 2022 \cite{LZ:2022lsv}, which for $M_{\rm LSP}$ scales approximately with the mass, with
\beq
 \left ( \frac{\Omega_{\rm LSP} h^2}{0.12} \right ) \sigma_{\rm SI} 
& \lsim &
\left (\frac{M_{\rm LSP}}{\mbox{1 TeV}} \right ) 2.8 \times 10^{-10}\>{\rm pb} 
.
\eeq
To illustrate the impact of this bound, I first consider models in which the wino and bino mass parameters at a renormalization scale $Q = 10$ TeV are taken in the ratio $M_2/M_1 = 1.8$, which is approximately the ratio implied by unification of gaugino masses at the scale of apparent gauge coupling unification above $Q=10^{16}$ GeV.
This implies that the higgsino-like LSP mixes more strongly with the bino, but
the wino contamination is still non-negligible. 
The masses of all squarks, sleptons, and the gluino are also taken to be 10 TeV. I fix the lightest Higgs scalar mass at $M_h = 125.1$ GeV, which is important because of its role in mediating the spin-independent LSP-nuclear cross-section. The other Higgs bosons are taken to be in the decoupling limit with masses also at 10 TeV. The physical neutralino and chargino masses and couplings are evaluated using the public code {\tt SOFTSUSY v4.1.12}
\cite{Allanach:2001kg}. 
It is useful to show constraints by plotting the double-scaled cross-section
\beq
\left (\frac{\mbox{1 TeV}}{M_{\rm LSP}} \right ) 
\left ( \frac{\Omega_{\rm LSP} h^2}{0.12} \right ) \sigma_{\rm SI} ,
\eeq
as the vertical axis, so that $\sigma_{\rm SI}$ limits (past, present, and future) are very nearly horizontal lines independent of the mass. 

\begin{figure}[!p]
\centering
\mbox{
\includegraphics[width=0.51\linewidth]{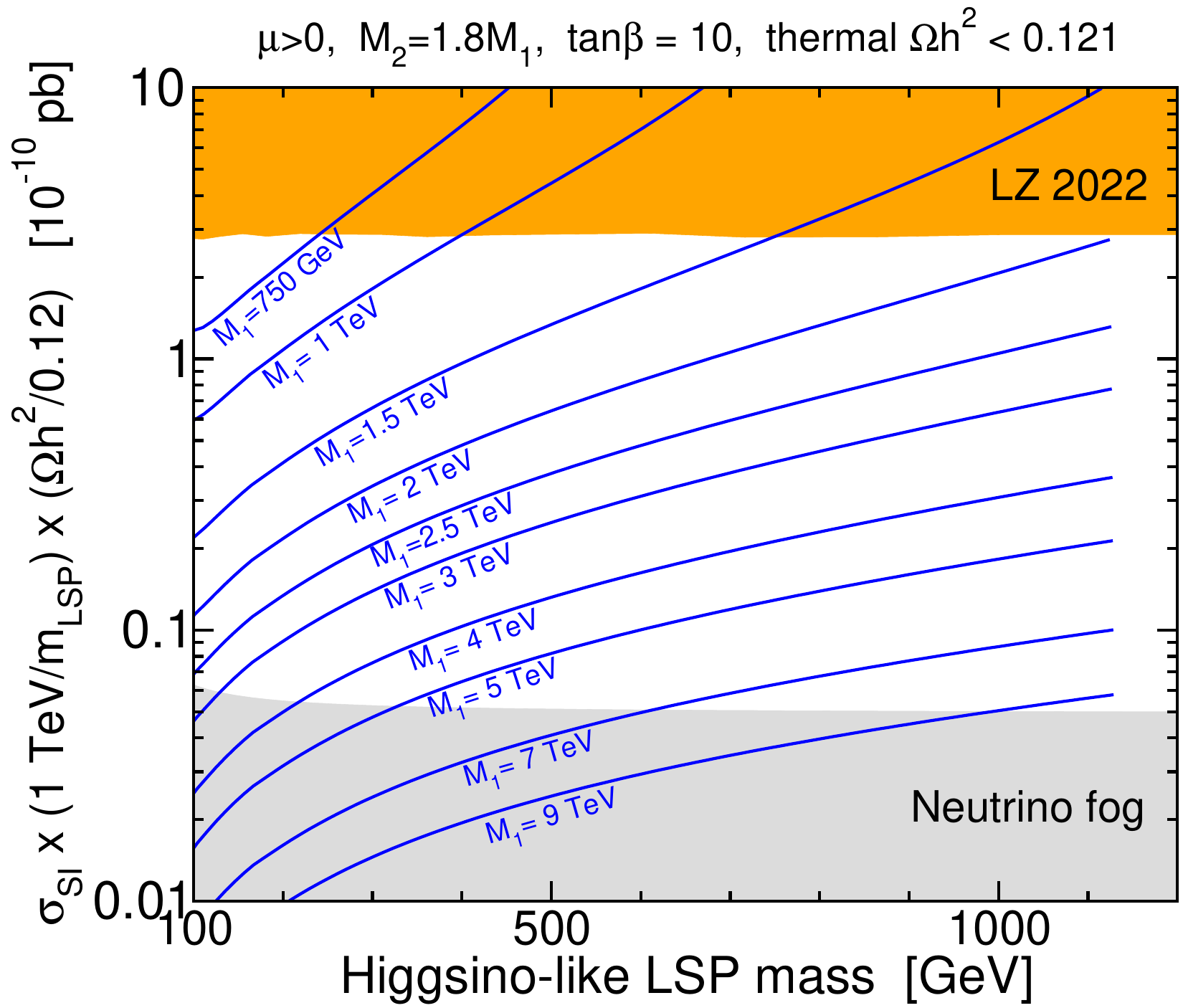}
\includegraphics[width=0.51\linewidth]{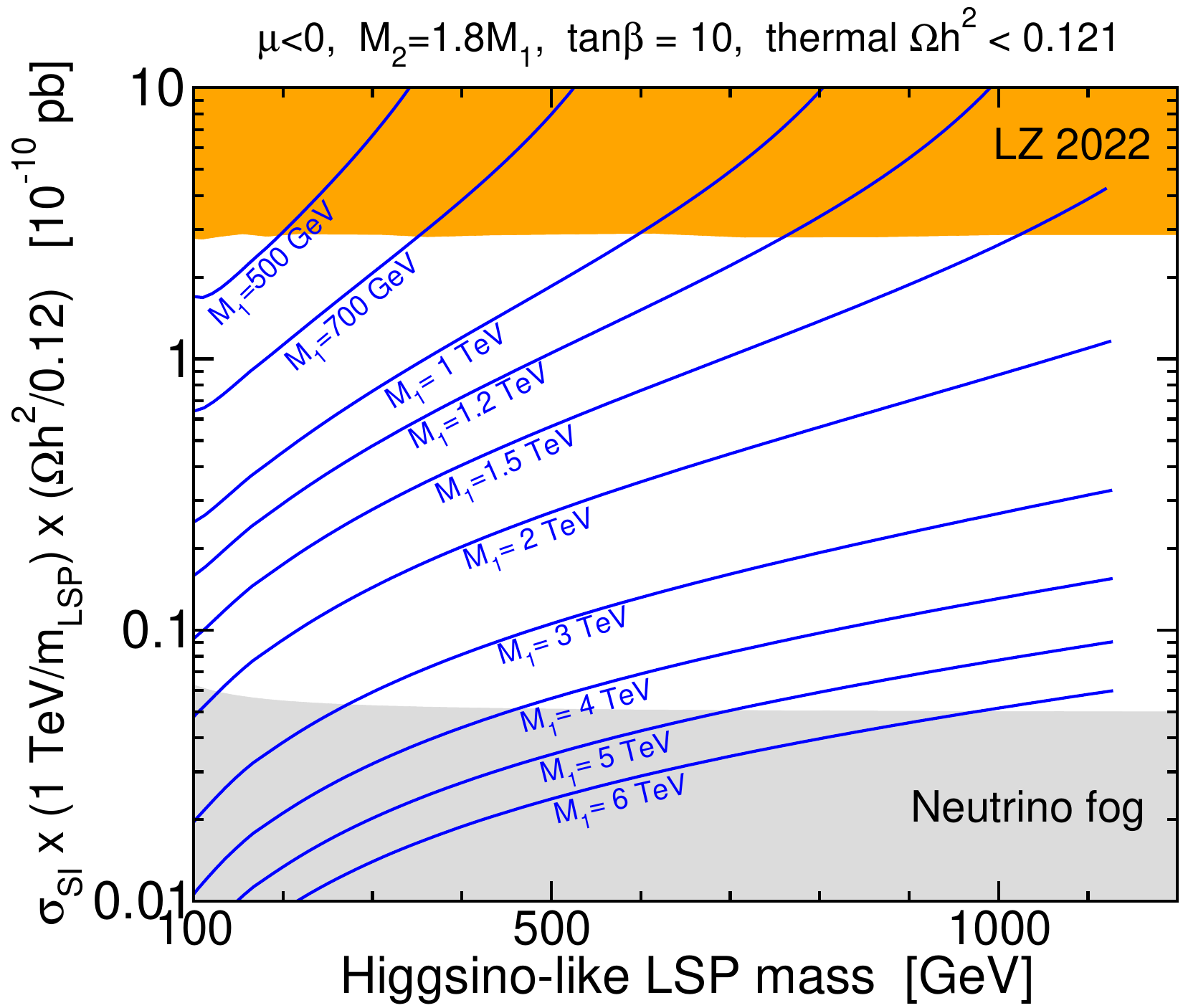}
}
\caption{\label{fig:M1mix}
The spin-independent LSP-nucleon cross-section, scaled by factors
of (1 TeV$/M_{\rm LSP}$) and $\Omega h^2/0.12$, as a function of the higgsino-like LSP mass $M_{\tilde N_1}$ (related and comparable to $|\mu|$), for various values of the Lagrangian parameter $M_1$, as
labeled. The wino mass parameter is taken to be $M_2 = 1.8 M_1$,
as motivated by models with gaugino mass unification, and $\tan\beta = 10$.
The masses of the squarks, sleptons, and gluino, and the heavy Higgs bosons 
$A, H, H^{\pm}$ are all set to 10 TeV, and $M_h = 125.1$ GeV is imposed.
The left panel shows results for $\mu>0$, and the right panel for $\mu < 0$.
The extent of the curves in the horizontal direction is set by the requirement $\Omega h^2 < 0.121$. The (orange) shaded band at top is the limit set by LZ 2022 \cite{LZ:2022lsv}. The (gray) shaded band at bottom is the expected neutrino fog level as defined in \cite{OHare:2021utq}.
}
\vspace{0.75cm}
\mbox{
\includegraphics[width=0.51\linewidth]{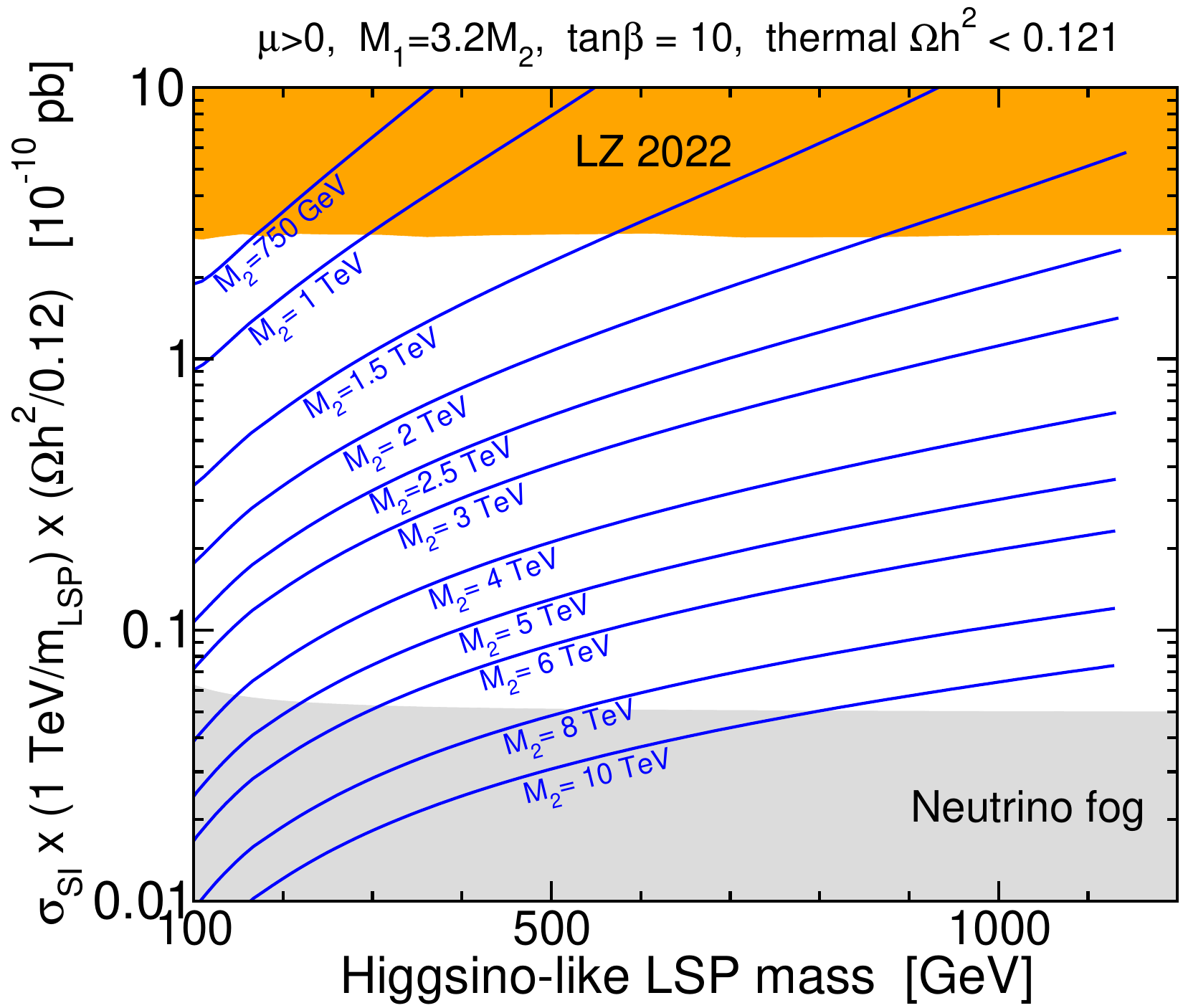}
\includegraphics[width=0.51\linewidth]{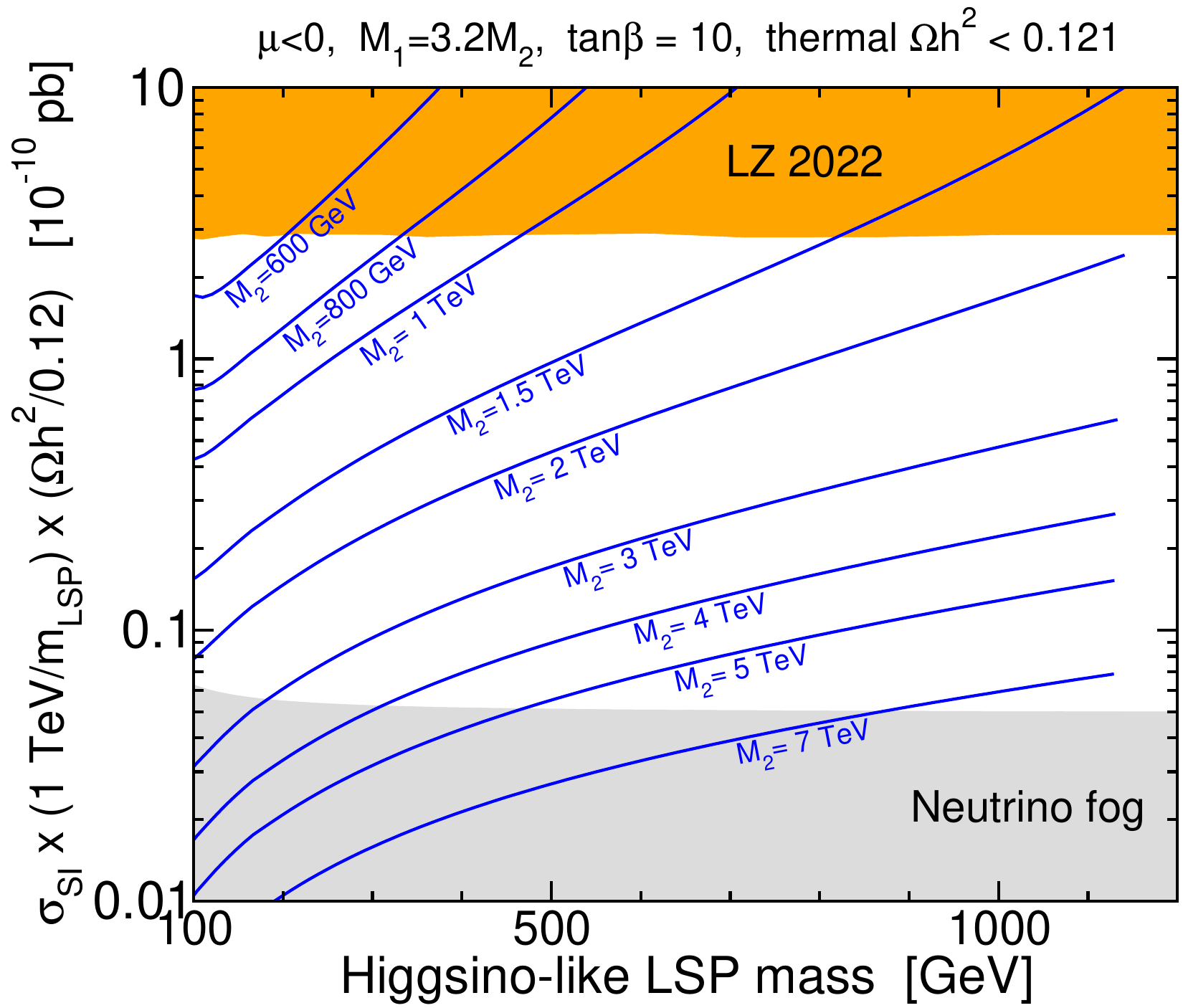}
}
\caption{\label{fig:M2mix}
The spin-independent LSP-nucleon cross-section, scaled by factors
of (1 TeV$/M_{\rm LSP}$) and $\Omega h^2/0.12$, as a function of the higgsino-like LSP mass $M_{\tilde N_1}$ (related and comparable to $|\mu|$), for various values of the Lagrangian parameter $M_2$, as
labeled. The bino mass parameter is taken to be $M_1 = 3.2 M_2$,
as motivated by AMSB models, and $\tan\beta = 10$.
The masses of the squarks, sleptons, and gluino, and the heavy Higgs bosons 
$A, H, H^{\pm}$ are all set to 10 TeV, and $M_h = 125.1$ GeV is imposed.
The left panel shows results for $\mu>0$, and the right panel for $\mu < 0$.
The extent of the curves in the horizontal direction is set by the requirement $\Omega h^2 < 0.121$. The (orange) shaded band at top is the limit set by LZ 2022 \cite{LZ:2022lsv}. The (gray) shaded band at bottom is the expected neutrino fog level as defined in \cite{OHare:2021utq}.
}
\end{figure}

The results are shown in Figure \ref{fig:M1mix} as a function of the LSP mass $M_{\tilde N_1}$, for curves corresponding to various fixed values of $M_1$, with $\tan\beta=10$, and positive $\mu$ in the left panel and negative $\mu$ in the right panel. The curves terminate slightly
above $M_{\tilde N_1} = 1.1$ TeV, due to the requirement
$\Omega_{\rm LSP} h^2 < 0.121$. The shaded (orange) region at top is the LZ 2022 excluded region.
These results show that the higgsino purity constraint requires $M_1$ to be greater than about 2 TeV for positive $\mu$, and greater than about 1.6 TeV for negative $\mu$, in the case that the higgsino is all of the dark matter. For lighter LSPs, $M_1$ can be considerably smaller, as shown, but there are still non-trivial lower bounds on $M_1$ despite the suppression from $\xi$. These bounds are reduced to well below 1 TeV if the higgsino mass is less than a few hundred GeV. It is a general feature that the lower bounds on $M_1$ are weaker for negative $\mu$, which can be understood from the expression
in eq.~(\ref{eq:yh}) for the LSP-Higgs coupling.

%%%%%%%%%%%%%%%%%%%%%%%%%%%%%%%%%%%%%%%%%%%%%%%%%%%%%%%%%%%%%%%%%%%%%%%%%%%%%%%%%
%%%%%%%%%%%%%%%%%%%%%%%%%%%%%%%%%%%%%%%%%%%%%%%%%%%%%%%%%%%%%%%%%%%%%%%%%%%%%%%%%

Also shown in Figure \ref{fig:M1mix} as the lower shaded (gray) region is the ``neutrino fog" level at which LSP discovery becomes problematic due to the background from scattering of astrophysical neutrinos \cite{Billard:2013qya}. This is not a completely solid floor, because it is subject to neutrino flux uncertainties and might be overcome to some limited extent by improved signal discrimination and statistics. Here, I use the definition for the neutrino fog level given in ref.~\cite{OHare:2021utq}. It can be seen that the unshaded region between the present LZ bound and the neutrino fog corresponds to a large range for $M_1$ that will be newly probed in future direct detection experiments, up to $M_1 = 9$ TeV for positive $\mu$ and $M_1 = 6$ TeV for negative $\mu$. This is far beyond what can be probed at the LHC, even with an energy upgrade. This reach is of course diminished for smaller LSP masses due to the $\xi$ suppression, but is seen to be still substantial.

The situation if the higgsinos instead mix mostly with winos is shown similarly
in Figure \ref{fig:M2mix}. To be specific, I choose $M_1 = 3.2 M_2$ (again at a renormalization scale of 10 TeV), inspired by the gaugino mass relations in models of anomaly-mediated supersymmetry breaking (AMSB) \cite{Randall:1998uk,Giudice:1998xp}. As one might expect from the fact that $g > g'$, the lower bounds on the wino mass are stronger than for mixing predominantly with the bino, reaching about 2.4 TeV for $\mu>0$ and 1.9 TeV for $\mu < 0$, in the case of higgsinos near 1.1 TeV that provide $\Omega h^2 = 0.12$. Conversely, the future 
direct detection reach in $M_2$ above the neutrino fog is greater, exceeding 10 TeV for $\mu>0$ and 7 TeV for $\mu < 0$.
  
\begin{figure}[!p]
\centering
\mbox{
\includegraphics[width=0.50\linewidth]{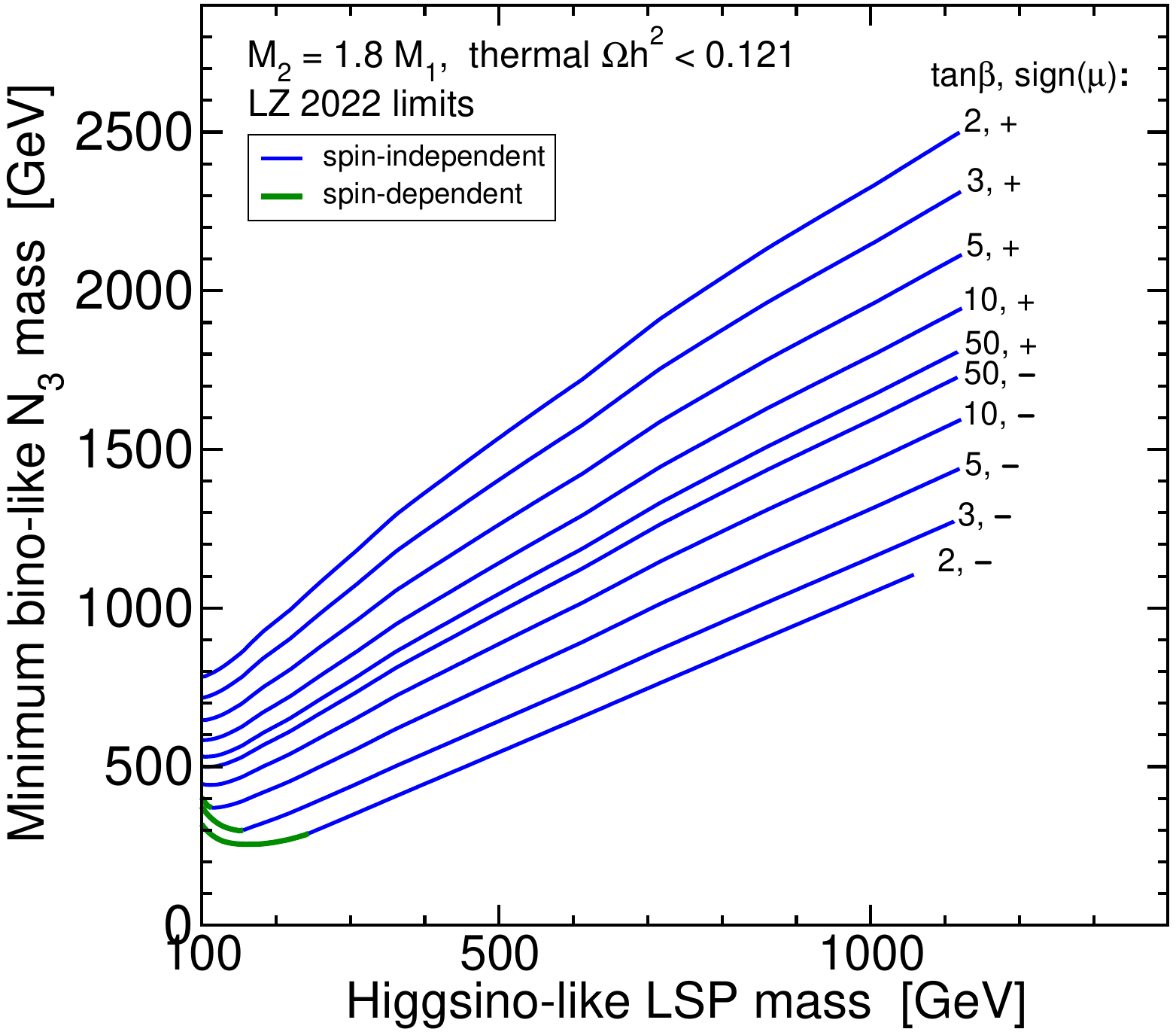}
\includegraphics[width=0.50\linewidth]{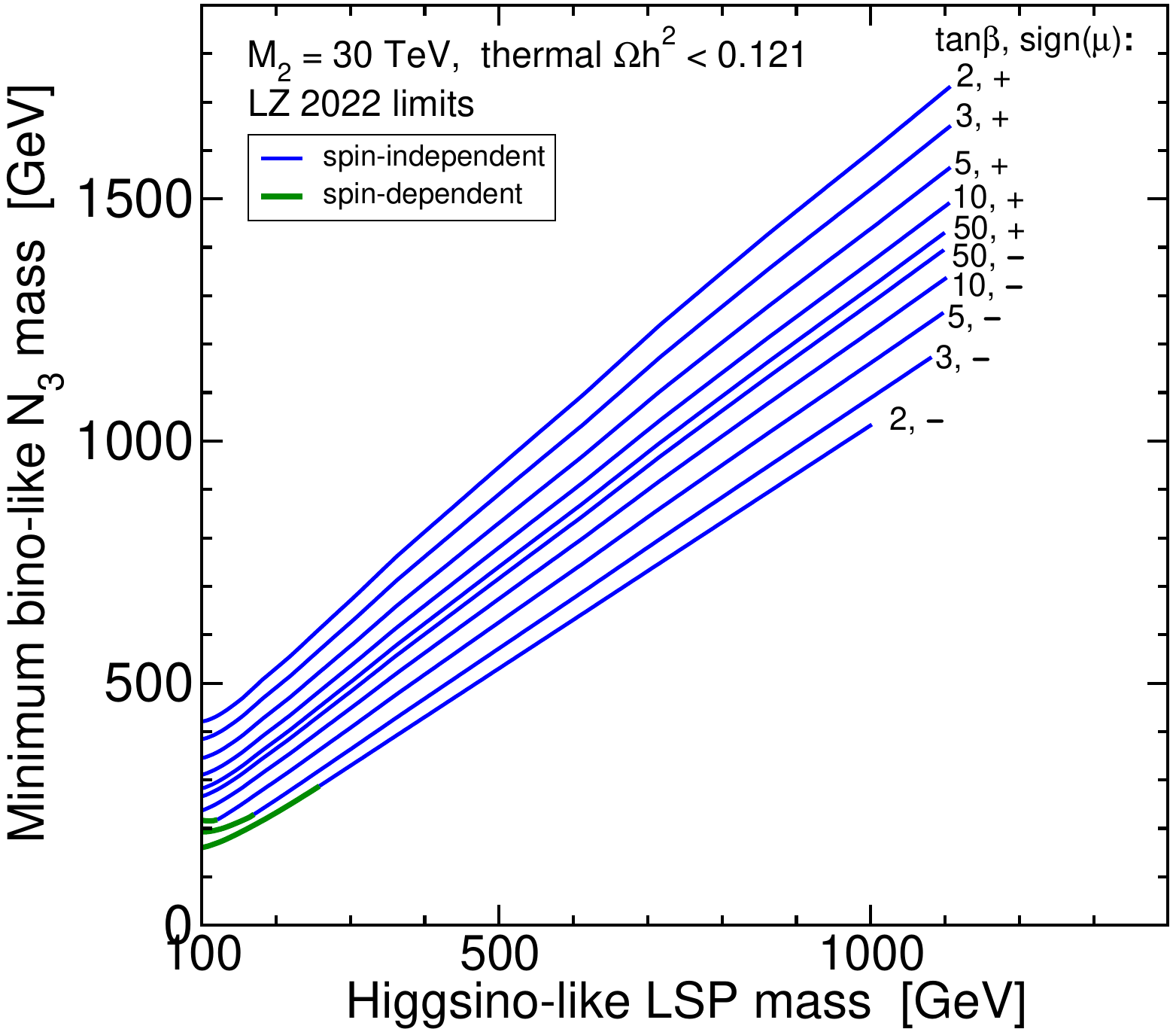}
}
\caption{\label{fig:contoursM1mix} 
The minimum bino-like neutralino mass $M_{\tilde N_3}$ (related to $M_1$) consistent with the LZ 2022 direct detection limits \cite{LZ:2022lsv},
as a function of the higgsino-like LSP mass $M_{\tilde N_1}$ (related to $|\mu|$). The LSP density is assumed to be the thermal freezeout prediction for
$\Omega h^2$, required to be less than $0.121$; this sets the maximum values of $M_{\tilde N_1}$. The different lines are the results for
various combinations of $\tan\beta$ and sign($\mu$), as labeled. 
The lower bound on the bino mass is set by 
the spin-independent
LSP-xenon nucleus cross-section, 
except for the thicker (green) portion of the curves at small $\tan\beta$ and negative $\mu$ and small LSP masses, where the spin-dependent
LSP-neutron cross-section sets the bound.
%For the thinner (blue) lines, the minimum is set by the spin-independent LSP-xenon nucleus cross-section limit.
%For the thicker (green) lines at negative $\mu$, small $\tan\beta$, and small masses, the minimum is instead set by the spin-dependent LSP-neutron cross-section. 
The left panel shows the results assuming $M_2 = 1.8 M_1$ 
as in models with gaugino mass unification, and the right panel instead takes the winos decoupled, with $M_2 = 30$ TeV. The
lightest Higgs boson mass is fixed to $M_h = 125.1$ GeV, and
all other superpartner and Higgs boson masses are set to 10 TeV. Note the different vertical scales.}
%\end{figure}
%\begin{figure}[!h]
%\centering
\vspace{0.35cm}
\mbox{
\includegraphics[width=0.50\linewidth]{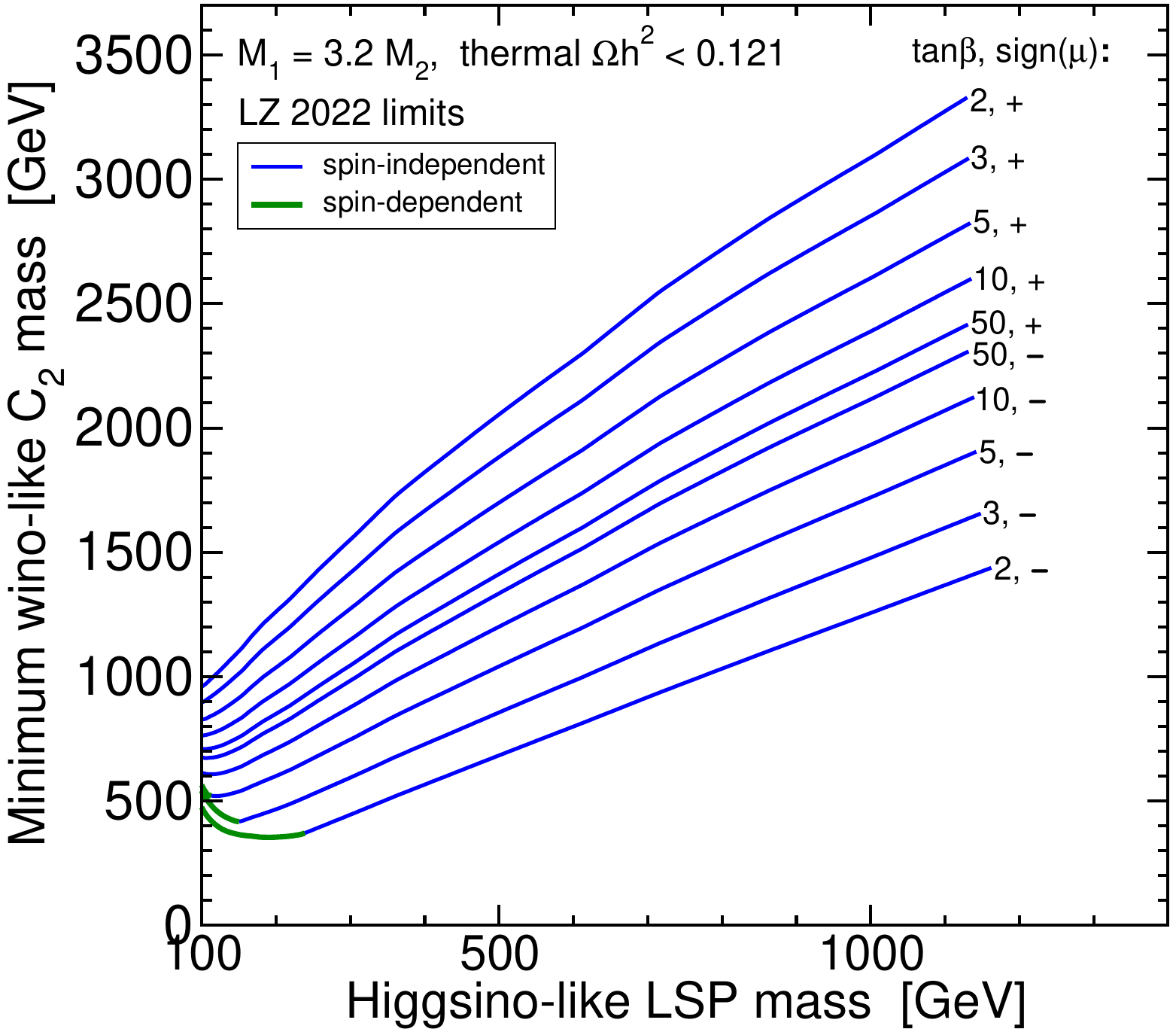}
\includegraphics[width=0.50\linewidth]{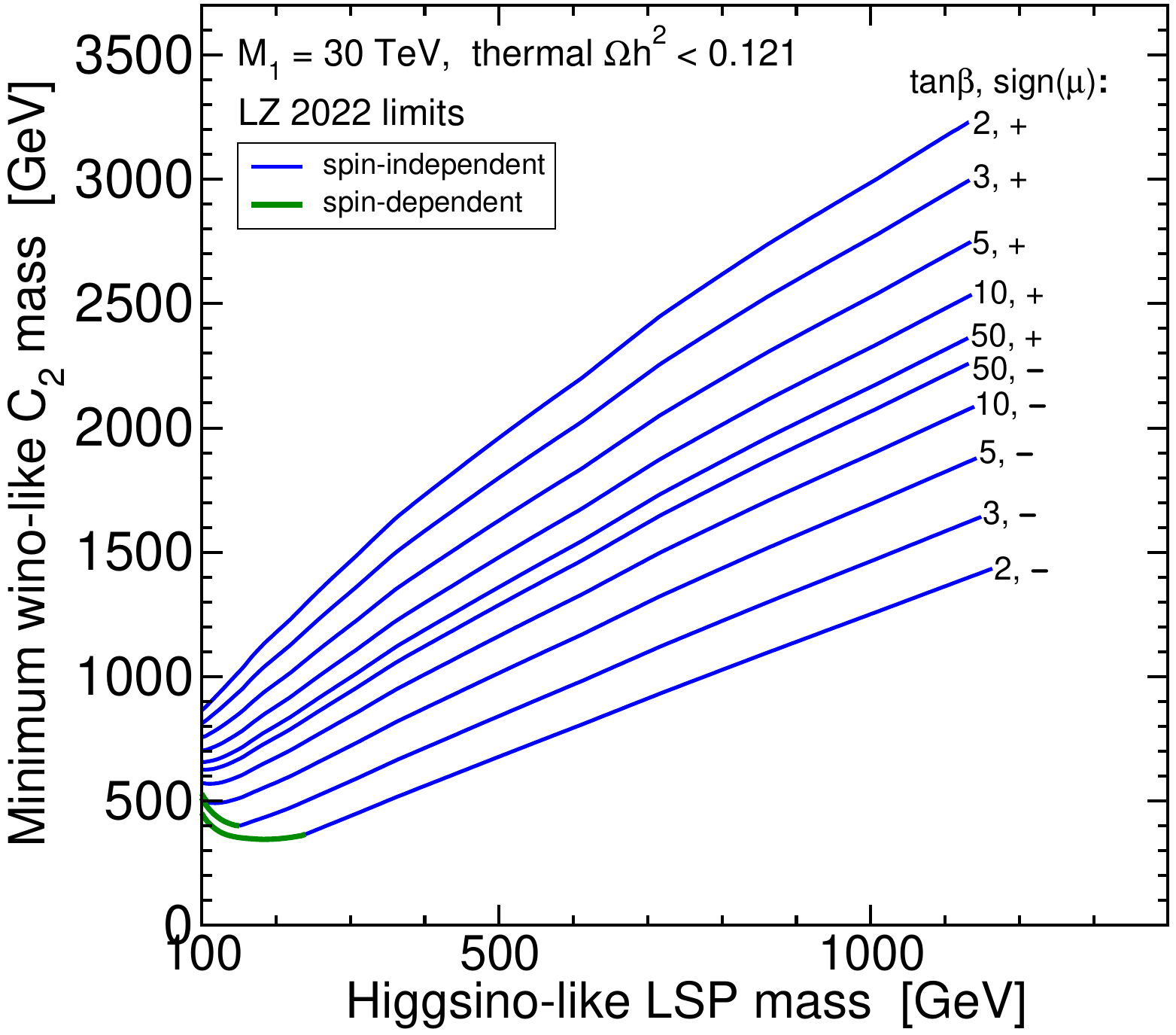}
}
\caption{\label{fig:contoursM2mix}
The minimum wino-like chargino mass $M_{\tilde C_2}$ (related to $M_2$) consistent with LZ 2022 direct detection limits \cite{LZ:2022lsv},
as a function of the higgsino-like LSP mass $M_{\tilde N_1}$ (related to  $|\mu|$). The LSP density is assumed to be the thermal freezeout prediction for
$\Omega h^2$, required to be less than $0.121$. The different lines are the results for
various combinations of $\tan\beta$ and sign($\mu$), as labeled. 
The lower bound on the wino mass is set by 
the spin-independent
LSP-xenon nucleus cross-section, except for 
the thicker (green) portion of the curves at small $\tan\beta$ and negative $\mu$ and small LSP masses where the spin-dependent
LSP-neutron cross-section sets the bound.
%For the thinner (blue) lines, the minimum is set by the spin-independent LSP-xenon nucleus cross-section limit. For the thicker (green) lines at negative $\mu$, small $\tan\beta$, and small masses, the minimum is instead set by the spin-dependent LSP-neutron cross-section. 
The left panel shows the results assuming $M_1 = 3.2 M_2$ 
as in AMSB models, and the right panel instead takes the bino decoupled, with $M_1 = 30$ TeV. The lightest Higgs boson mass is fixed to $M_h = 125.1$ GeV, and
all other superpartner and Higgs boson masses are set to 10 TeV.}
\end{figure}

The results in Figures \ref{fig:M1mix} and \ref{fig:M2mix} assumed $\tan\beta=10$ for simplicity, but there is a rather strong dependence on $\tan\beta$. To show how this goes, Figure \ref{fig:contoursM1mix} gives the minimum allowed bino-like neutralino mass $M_{\tilde N_3}$ (which is closely correlated with the bino-mass parameter $M_1$), as a function of the higgsino-like LSP mass $M_{\tilde N_1}$, with separate contours for various values of $\tan\beta$ and the sign of $\mu$. Here, I consider two cases. The left panel shows
the results assuming the gaugino-mass unification condition $M_2 = 1.8 M_1$, while
the right panel shows the results assuming that the wino is thoroughly decoupled
at $M_2 = 30$ TeV.
In both cases, the squark, slepton, gluino, and other Higgs masses are taken to be nearly decoupled at 10 TeV, while the lightest Higgs mass is always fixed at 125.1 GeV. For most parameters, the lower bound on $M_2$ is set by the
LZ limit on the spin-independent cross-section. However, as we saw in eq.~(\ref{eq:yh}),
the LSP-Higgs coupling, and therefore $\sigma_{\rm SI}$, are relatively
suppressed for negative $\mu$ and small $\tan\beta$. Therefore, in some cases with
$\tan\beta < 5$ and $\mu<0$ it turns out that the strongest bound for small higgsino masses is obtained instead from the LZ limit on the spin-dependent LSP-neutron cross-section \cite{LZ:2022lsv}, which to a good approximation is 
\beq
\left ( \frac{\Omega_{\rm LSP} h^2}{0.12} \right ) \sigma^n_{\rm SD} 
&\lsim&
\left (\frac{M_{\rm LSP}}{\mbox{1 TeV}} \right ) 4.9 \times 10^{-5}\>{\rm pb} 
\eeq 
in the relevant mass range above 100 GeV.
The cases where the spin-dependent neutron cross-section set the bound are shown in Figure \ref{fig:contoursM1mix} as the heavier (green) parts of the curves. 
Note that the small contamination of the heavier wino mixing in the LSP in the left panel results in a stronger lower bound on $M_1$ compared to the case in the right panel in which the wino is completely decoupled. In general, the lower bounds on $M_1$
are weakest when $\tan\beta$ is small and $\mu$ is negative, and strongest bounds when 
$\tan\beta$ is small and $\mu$ is positive, reflecting the behavior found for $y_h$ in eq.~(\ref{eq:yh}). For a thermal dark matter higgsino-like LSP with mass near 1.1 TeV so that $\xi=1$, the bino mass is only required to be slightly larger than $|\mu|$ when $\tan\beta=2$ and $\mu<0$, but must be over 1.7 TeV for $\tan\beta=2$ and $\mu>0$, and at least 2.5 TeV if one assumes gaugino mass unification.

For the case that the higgsino-like LSP mixes predominantly with the wino, 
the lower bounds on $M_{\tilde C_2}$ (closely related to $M_2$) are given in Figure \ref{fig:contoursM2mix}, again for various combinations of $\tan\beta$ and sign$(\mu)$. The left panel shows the AMSB-inspired case of $M_1 = 3.2 M_2$, while the right panel shows the case that the bino is essentially completely decoupled, with $M_1$ fixed at 30 TeV. As can be seen from the figure, the lower bound on $M_2$ is only slightly weakened by completely decoupling the bino, compared to the AMSB case in the left panel. These bounds are stronger than for predominantly bino mixing, especially
for positive $\mu$. One can conclude that for higgsino-like thermal dark matter with $\xi=1$ and mixing predominantly with the wino, $M_2$ can be as small as less than 1.5 TeV if $\tan\beta=2$ and $\mu$ is negative, but the lower bounds exceed 3 TeV when $\mu$ is positive. The more relevant scenario for a possible future discovery at the LHC obviously occurs when $|\mu|$ is less than a few hundred GeV, with the most optimistic cases arising for negative $\mu$.

%%%%%%%%%%%%%%%%%%%%%%%%%%%%%%%%%%%%%%%%%%%%%%%%%%%%%%%%%%%%%%%%%%%%%%%%%%%%%%%%%
%%%%%%%%%%%%%%%%%%%%%%%%%%%%%%%%%%%%%%%%%%%%%%%%%%%%%%%%%%%%%%%%%%%%%%%%%%%%%%%%%

\clearpage
\baselineskip=15.6pt

\section{Bounds on higgsino mass splittings\label{sec:deltambounds}}
\setcounter{equation}{0}
\setcounter{figure}{0}
\setcounter{table}{0} 
\setcounter{footnote}{1}

The purity constraints on higgsino-like LSPs as part of the dark matter are also reflected in upper bounds on the mass splittings. This is particularly important for LHC and future hadron collider searches, which typically rely on either having mass splittings large enough for the decay products of the heavier higgsinos to pass cuts, or small enough so that quasi-stable higgsinos can leave disappearing tracks or displaced vertices. Between those two cases is a regime that is especially difficult to probe at hadron colliders \cite{Baer:2011ec}, and for which the current exclusion reach is very modest. In fact, the present official exclusion reach from the LHC experimental collaborations is minimal for the LSP mass above 100 GeV when $\Delta M_0$ is between about 1 and 3 GeV, but it has been suggested \cite{Agin:2023yoq} that this gap might be closed with existing monojet searches \cite{ATLAS:2021kxv,CMS:2021far}. For $\Delta M_0$ less than about 40 GeV, the expected exclusion reach from soft leptons is presently limited to higgsino masses below about 200 GeV, and the observed exclusion is less. As we will now see, in the case of decoupled MSSM scalars (other than the 125 GeV Higgs boson), the mass splittings for higgsinos are required by LZ 2022 to be not too large, and are in a difficult but interesting region for LHC searches, if the dark matter abundance is set by thermal freezeout.

\begin{figure}[!p]
\centering
\mbox{
\includegraphics[width=0.52\linewidth]{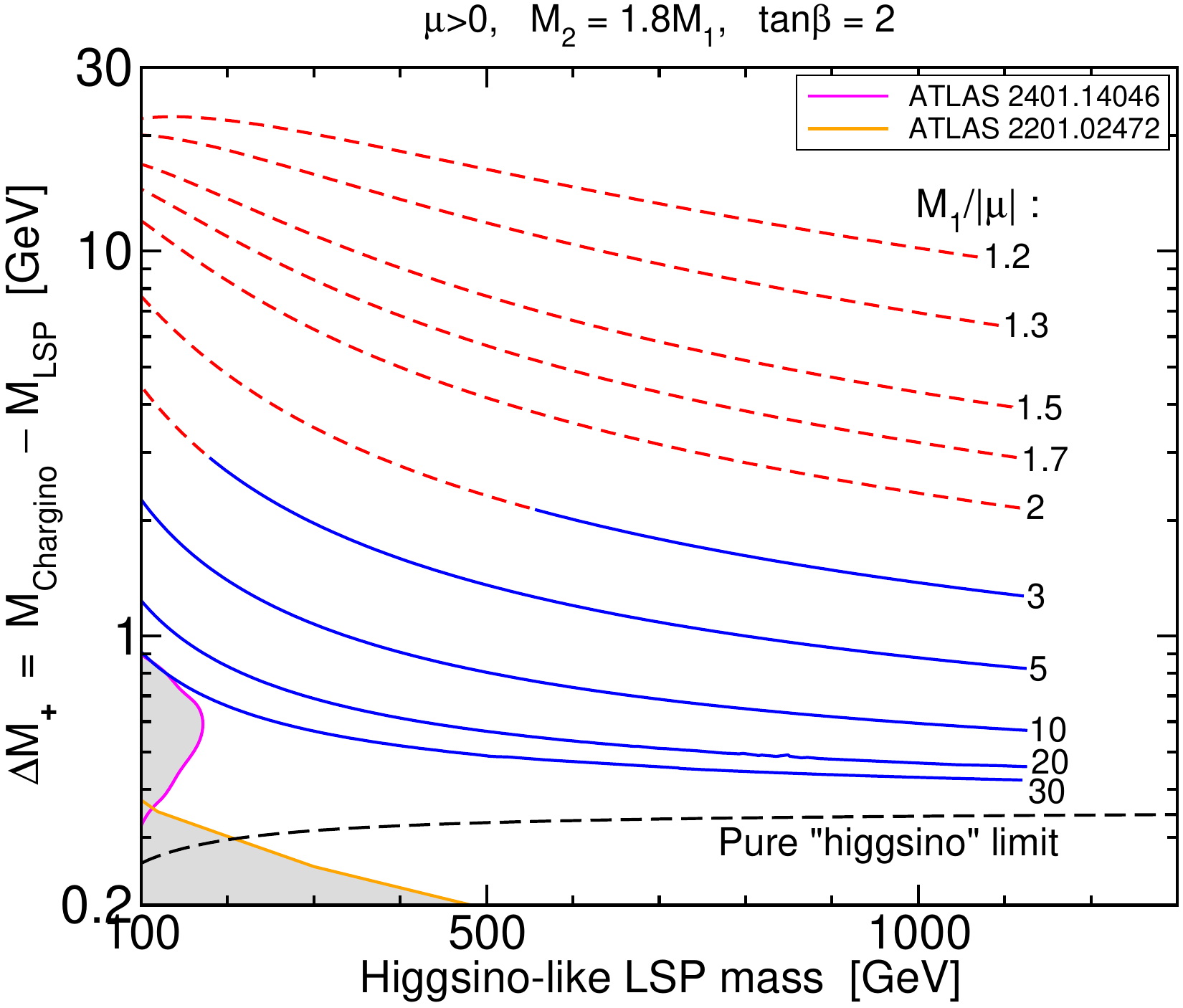}
\includegraphics[width=0.52\linewidth]{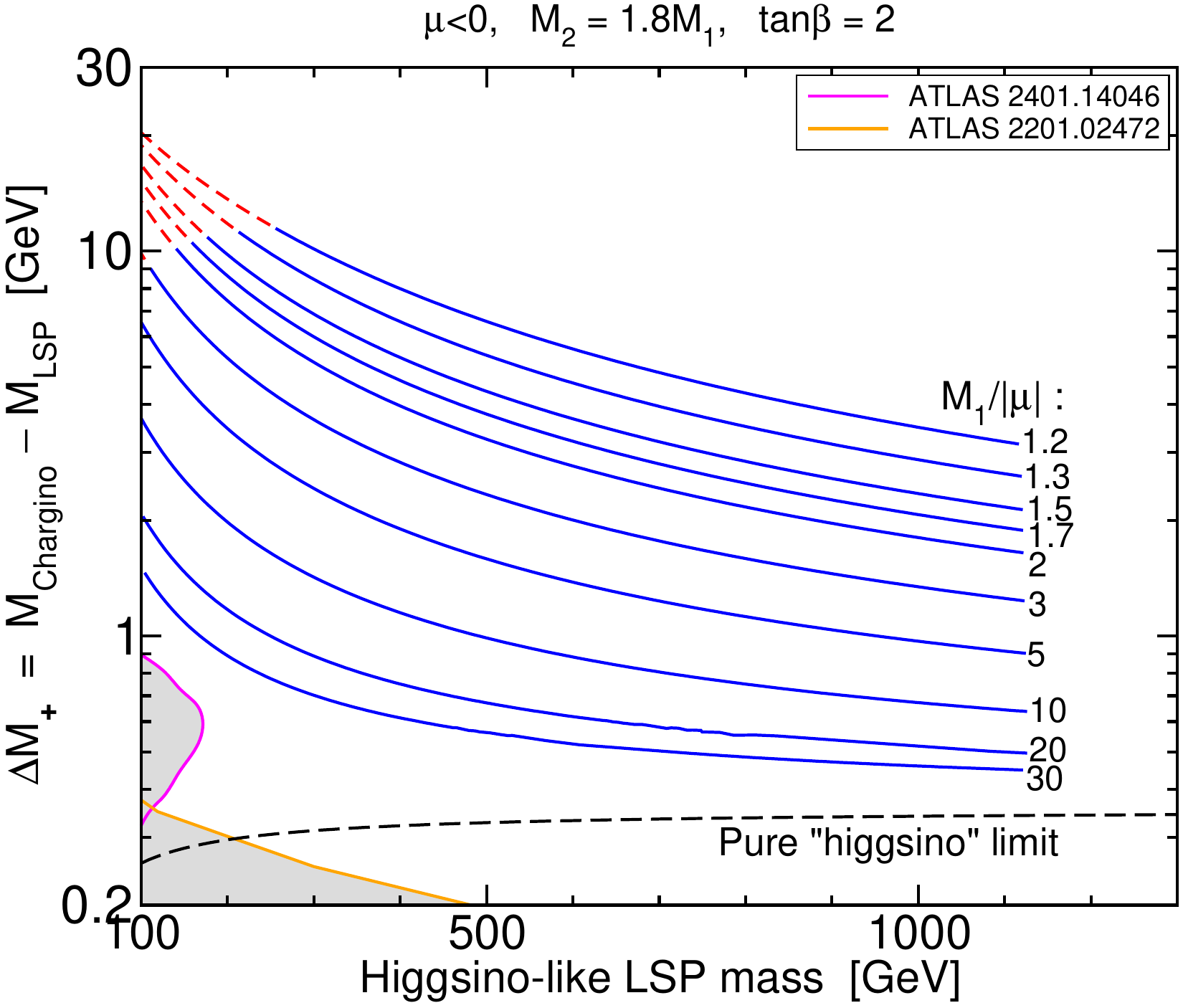}
}
\mbox{
\includegraphics[width=0.52\linewidth]{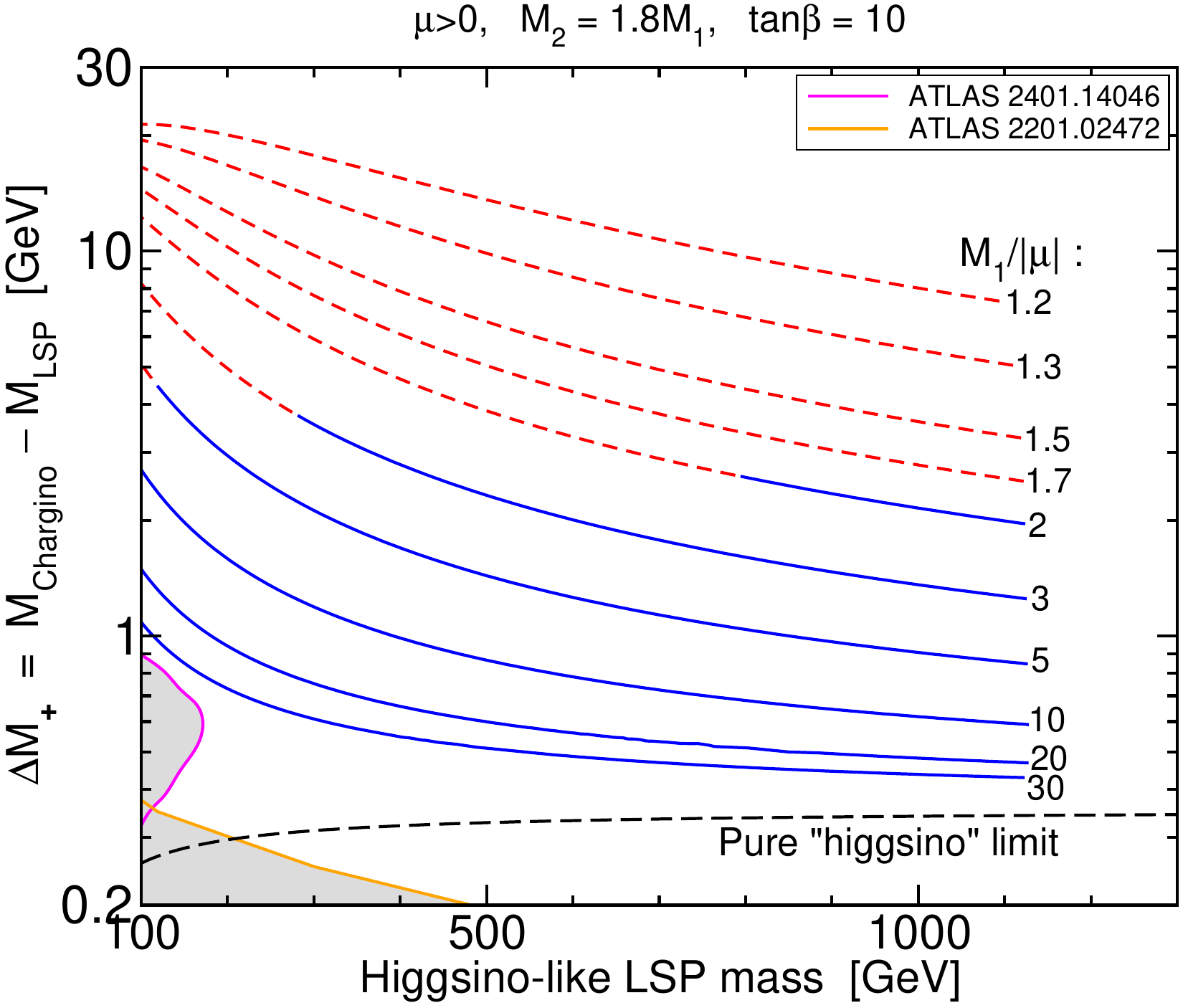}
\includegraphics[width=0.52\linewidth]{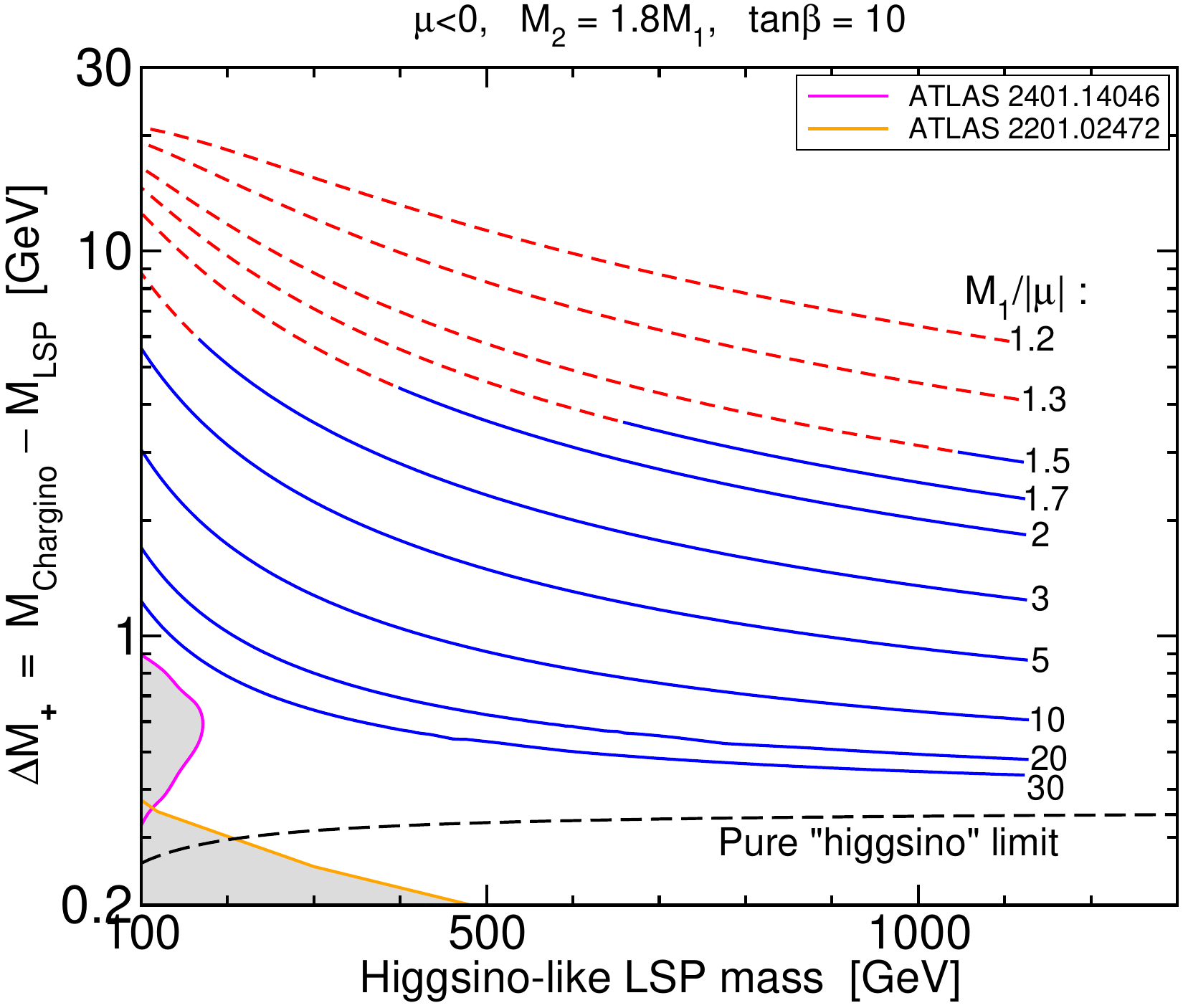}
}
%\mbox{
%\includegraphics[width=0.55\linewidth]{DeltaMplus_posmu_50.pdf}
%\includegraphics[width=0.55\linewidth]{DeltaMplus_negmu_50.pdf}
%}
\caption{\label{fig:deltaMp}
The mass splitting $\Delta M_+ \equiv M_{\tilde C_1} - M_{\tilde N_1}$ for mostly higgsino-like
charginos and neutralinos, as a function of the LSP mass $M_{\tilde N_1}$, for various
ratios $M_1/|\mu| = 1.2, 1.3, 1.5, 1.7, 2, 3, 5, 10, 20,$ and $30$, as the labeled curves.
In each case, the higgsino mixes more strongly with the bino than the wino, with $M_2 = 1.8 M_1$
as motivated by models with gaugino mass unification.
The dark matter density $\Omega h^2$ is assumed to be the thermal freezeout result, and is required to be
$<0.121$; this sets the horizontal range of the curves. 
The lightest Higgs mass is fixed at 125.1 GeV. The squark, slepton, gluino, and heavy Higgs boson $A, H, H^{\pm}$ masses are set to 10 TeV.
The solid (blue) portions of the curves satisfy the LZ 2022 limits \cite{LZ:2022lsv} on both spin-independent and spin-dependent cross-sections, while the dashed (red) portions do not.
The four panels show results for $\tan\beta=2,10$ and sign$(\mu) = \pm$.
In each panel, the shaded (gray) regions at lower left are constraints from ATLAS searches
for low-momentum mildly displaced tracks \cite{ATLAS:2024umc} and disappearing tracks \cite{ATLAS:2022rme}. There is a very similar result to the latter from CMS \cite{CMS:2023mny}. 
Also shown is the mass-splitting prediction \cite{Thomas:1998wy} for pure ``higgsinos", in the idealized case that the gauginos are taken to be absent or completely decoupled.}
\end{figure}

Figure \ref{fig:deltaMp} shows the higgsino-like chargino-LSP mass splitting $\Delta M_+$, as a function of the LSP mass, for models in which the neutral higgsinos mix with a heavier bino. Here, it is assumed that there is approximate gaugino mass unification at the GUT scale, so that $M_2 = 1.8 M_1$ at a renormalization scale $Q=10$ TeV, so that the wino contamination is smaller but non-negligible. The left panels have positive $\mu$, and the right panels have negative $\mu$,
and the upper panels have $\tan\beta=2$ while the lower panels have $\tan\beta=10$. As before, the squarks, sleptons, gluino, and the heavy Higgs scalars $A, H, H^{\pm}$ have masses set to 10 TeV. The different curves show results for constant ratios $M_1/|\mu|$ ranging from 1.2 to 30, as labeled. The dashed parts of the curves have $\sigma_{\rm SI}$ in excess of the LZ 2022 bound, taking into account the fact that the dark matter density is lower than the value implied by the Planck experiment observations ($\xi<1$). This means that only the solid parts of the curves are nominally allowed, which accordingly limits the range of $\Delta M_+$. 

%%%%%%%%%%%%%%%%%%%%%%%%%%%%%%%%%%%%%%%%%%%%%%%%%%%%%%%%%%%%%%%%%%%%%%%%%%%%%%%%%
%%%%%%%%%%%%%%%%%%%%%%%%%%%%%%%%%%%%%%%%%%%%%%%%%%%%%%%%%%%%%%%%%%%%%%%%%%%%%%%%%

Also shown in Figure \ref{fig:deltaMp} are the current exclusion region from disappearing
tracks from ATLAS and CMS, as well as a recent exclusion from mildly displaced tracks,
obtained by ATLAS \cite{ATLAS:2024umc} following the suggestion of \cite{Fukuda:2019kbp}.
Although these searches set non-trivial limits on pure higgsino simplified models, the
disappearing track searches typically do not affect the parameter space of actual MSSM higgsinos unless the hierarchy $M_1/|\mu|$ is considerably larger than the highest choice of 30 shown in the plot. This holds more generally except for a small region in parameter space where one can tune the tree-level mass splitting $\Delta M_+$ in eq.~(\ref{eq:deltaMp}) to very small values. 
The ATLAS mildly displaced track search probes more moderate gaugino/higgsino mass hierarchies, but for the choices shown this search only grazes the boundary of the region obtained with $M_1/|\mu| = 30$ for $\tan\beta=2$ and positive $\mu$.

\begin{figure}[!t]
\mbox{
\includegraphics[width=0.52\linewidth]{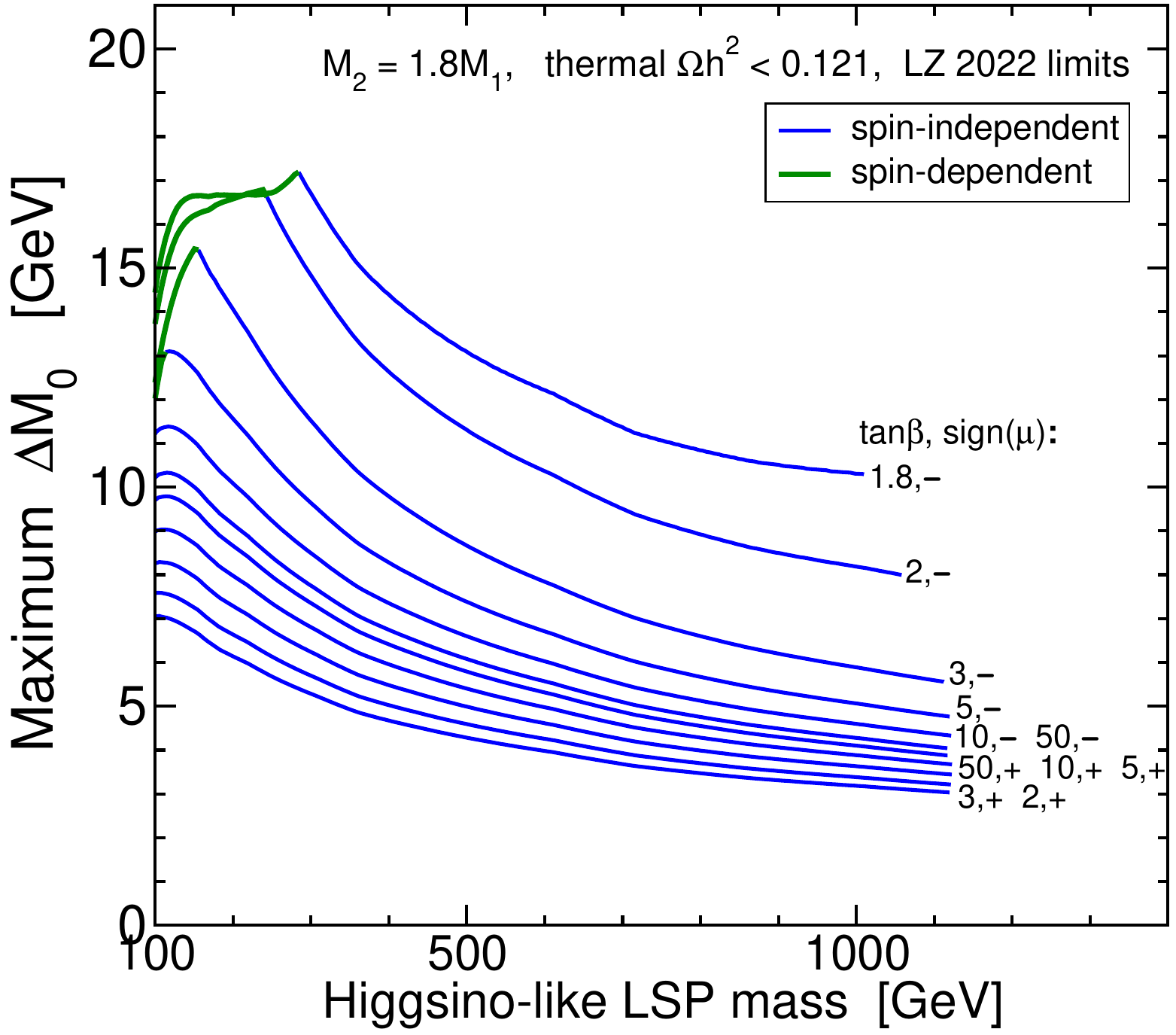}
\includegraphics[width=0.52\linewidth]{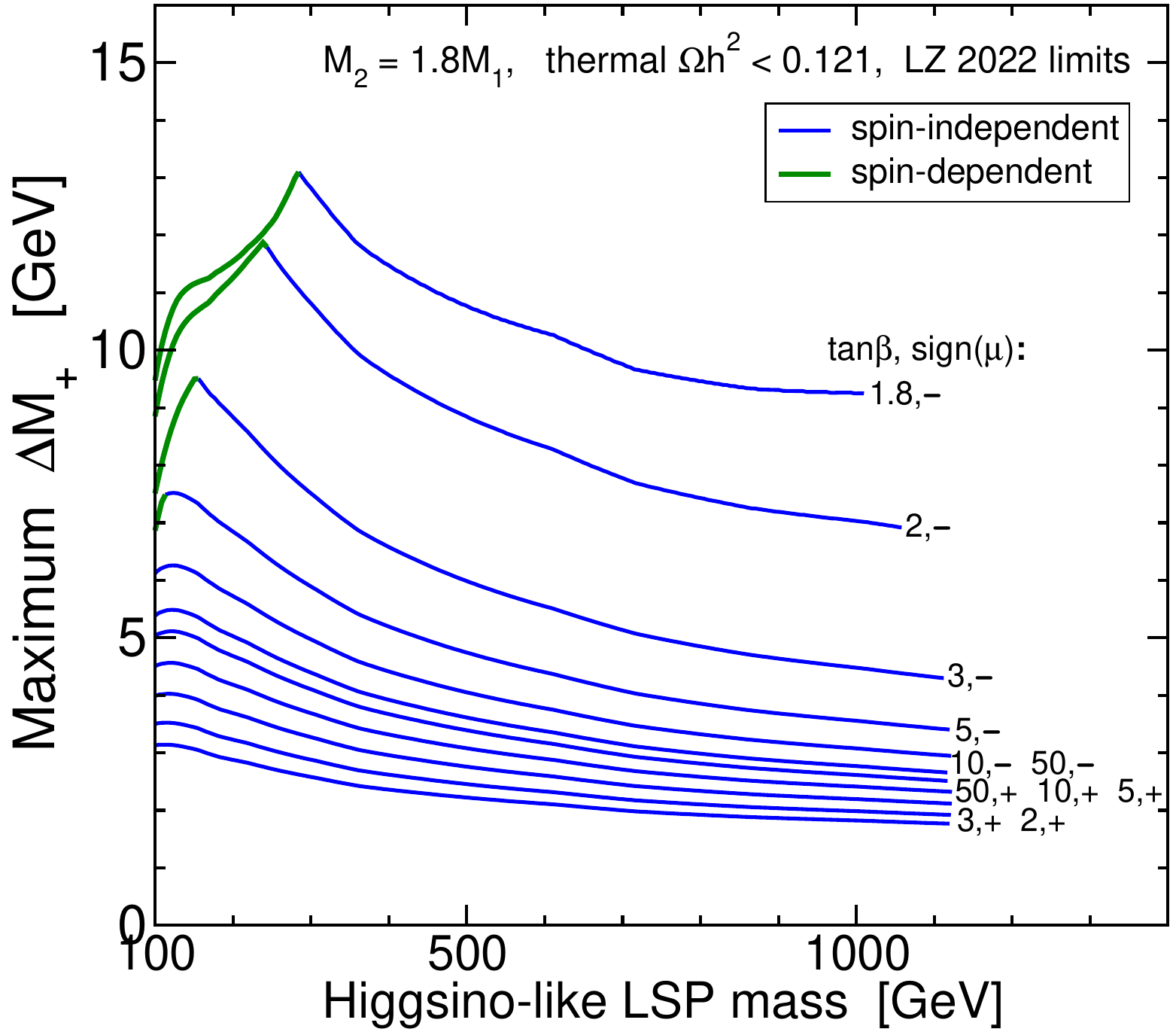}
}
\caption{\label{fig:maxDeltam}
The maximum higgsino-like mass splittings $\Delta M_0 \equiv M_{\tilde N_2}- M_{\tilde N_1}$ (left panel)
and $\Delta M_+ \equiv M_{\tilde C_1}- M_{\tilde N_1}$ (right panel), as functions of
the higgsino-like LSP mass $M_{\tilde N_1}$, for models that satisfy the direct detection
limits from LZ 2022 \cite{LZ:2022lsv}, for various combinations of $\tan\beta$ and 
sign$(\mu)$ as labeled. The gaugino masses are taken to satisfy $M_2 = 1.8 M_1$ at $Q=10$ TeV, as motivated by models with gaugino mass unification, so that the small mixing of the higgsino is mostly with the bino, but with a non-negligible wino contamination. The lightest Higgs mass is fixed at 125.1 GeV. The squark, slepton, gluino, and heavy Higgs boson $A, H, H^{\pm}$ masses are set to 10 TeV.  The dark matter density is assumed to be the thermal freezeout result, and is required to satisfy $\Omega h^2 < 0.121$; this sets the horizontal extent of the curves. The upper bounds on the mass splittings are set by 
the spin-independent
LSP-xenon nucleus cross-section, except for the thicker (green) portions of the curves at small $\tan\beta$ and negative $\mu$ and small LSP masses, where it is the spin-dependent
LSP-neutron cross-section that sets the bounds.}
\end{figure}

For larger mass splittings, the region allowed by LZ 2022 is significantly constrained. This is shown in more detail in Figure \ref{fig:maxDeltam},
which gives the maximum mass splittings $\Delta M_0$ (left panel) and $\Delta M_+$ (right panel), as a function of the LSP mass. Different curves show the results for various
combinations of $\tan\beta$ and sign$(\mu)$. As remarked several times above, the LZ 2022
$\sigma_{\rm SI}$ constraints are considerably weakened for negative $\mu$ and small $\tan\beta$. In this case, for small enough LSP mass, it is actually the spin-dependent LSP-neutron cross-section $\sigma_{\rm SD}^n$ that sets the bound. For $\tan\beta = 1.8$, the nominal bounds are about $\Delta M_0 < 17$ GeV and $\Delta M_+ < 13$ GeV, although in the present experimentally critical regime
of LSP masses less than 300 GeV, the bound on $\Delta M_+$ is stronger.  One can also read off the plot the maximum mass splittings for higgsino-like
particles with $M_{\tilde N_1}$ near 1.1 TeV, corresponding to $\Omega h^2 = 0.12$.
The bounds shown reflect the 90\% confidence level LZ 2022 limits, but it should be noted that there are significant uncertainties from nuclear matrix elements and from the local dark matter profile, so these bounds might be relaxed somewhat. 

%%%%%%%%%%%%%%%%%%%%%%%%%%%%%%%%%%%%%%%%%%%%%%%%%%%%%%%%%%%%%%%%%%%%%%%%%%%%%%%%%
%%%%%%%%%%%%%%%%%%%%%%%%%%%%%%%%%%%%%%%%%%%%%%%%%%%%%%%%%%%%%%%%%%%%%%%%%%%%%%%%%

ATLAS and CMS have presented results for searches sensitive to higgsinos with moderate mass splittings, using signal regions involving two or three soft leptons, which could arise from the decays $\tilde N_2 \rightarrow Z^{(*)} \tilde N_1$ and $\tilde C_1 \rightarrow W^{(*)} \tilde N_1$. For both experiments, the observed exclusions are weaker than expected, indicating a mild excess with respect to the estimated backgrounds. It has also recently been pointed out \cite{Agin:2023yoq} that monojet searches 
\cite{ATLAS:2021kxv,CMS:2021far}
may also have an excess in a kinematically consistent region. As a matter of opinion, I think that it may be too early to make more detailed fits of supersymmetric model parameters to such a mild excess. Doing so would be made more difficult because the 
published experimental search results make kinematic assumptions that need not hold in real supersymmetric models. In particular, refs.~\cite{ATLAS:2021moa} and \cite{CMS:2021edw} show search limits for higgsinos based on the assumption $\Delta M_+ = 0.5 \Delta M_0$, while ref.~\cite{CMS:2023qhl} shows results based on $\Delta M_+ = \Delta M_0$. As noted above, these are not expected to hold accurately for light higgsinos with significant mass splittings.
The efficiencies of searches in the soft lepton regime might be quite sensitive to those kinematic assumptions in ways that may make it difficult to extrapolate reliably. However, it seems worthwhile to make some comments regarding the possibility of accommodating the possible excess with the light higgsino models of the type considered in this paper.

\begin{figure}[!t]
\mbox{
\includegraphics[width=0.52\linewidth]{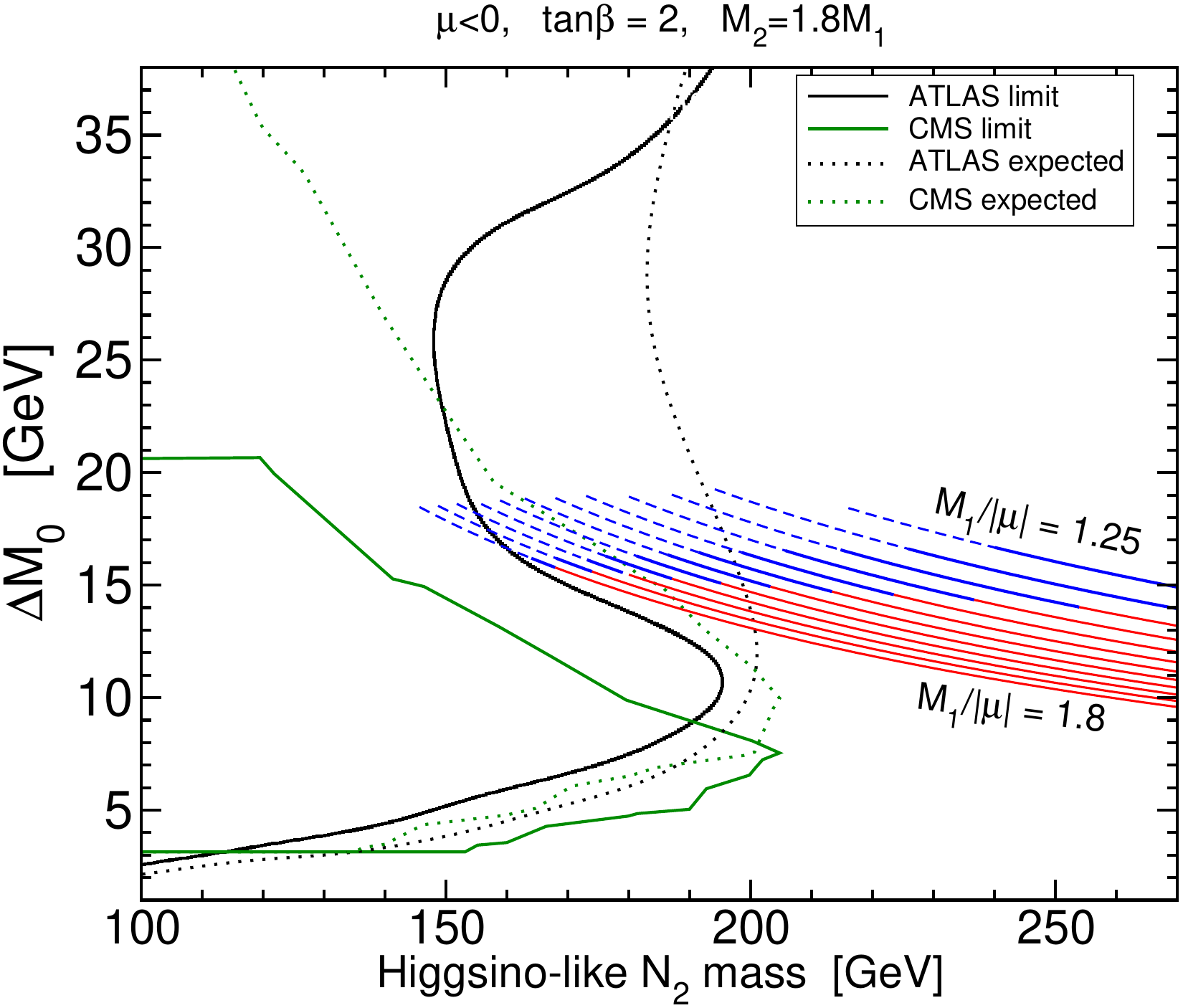}
\includegraphics[width=0.52\linewidth]{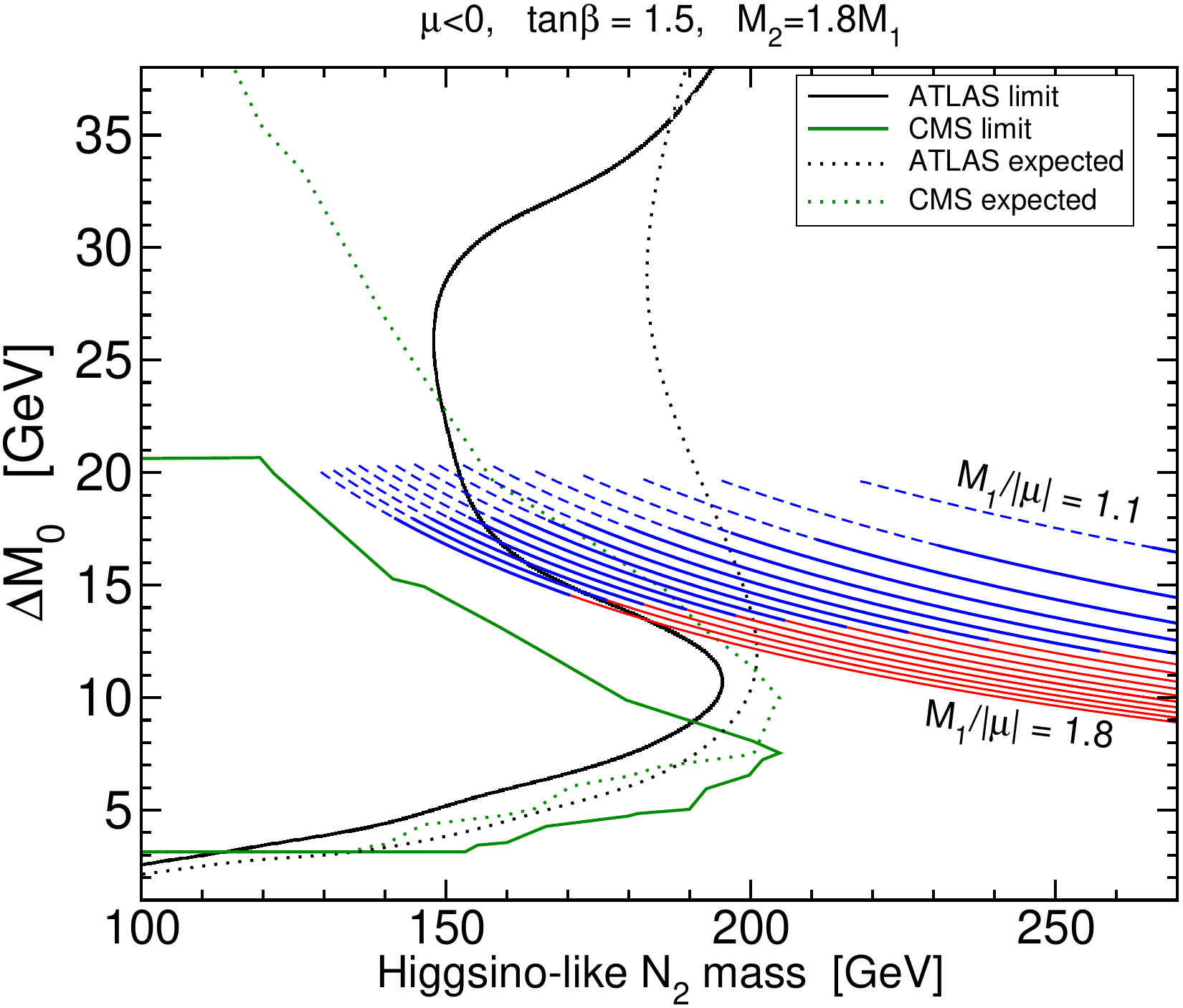}
}
\caption{\label{fig:DeltaM0}
The higgsino-like mass splitting $\Delta M_0 = M_{\tilde N_2} - M_{\tilde N_1}$,
as a function of $M_{\tilde N_2}$, for models with $\mu<0$ and $\tan\beta=2$ (left panel)
and $\tan\beta=1.5$ (right panel). 
Also shown are the ATLAS \cite{ATLAS:2021moa} and CMS \cite{CMS:2021edw} higgsino soft lepton search limits (solid curves) and expected limits (dotted curves) assuming $M_{\tilde C_1} = 0.5 (M_{\tilde N_1} + M_{\tilde N_2})$.
In both panels, $M_2 = 1.8 M_1$ at a renormalization scale $Q=10$ TeV,
as motivated by models with gaugino mass unification at high scales.
The model lines take the ratio $M_1/|\mu|$ in increments of 0.05, from 1.25 to 1.8 (left panel) and 1.1 to 1.8 (right panel). The solid portions of these lines are allowed by the nominal LZ 2022 \cite{LZ:2022lsv} direct detection searches, while the dashed portions relax the spin-dependent LSP-neutron cross-section limits by a factor of 2, in recognition of the significant uncertainties in neutron matrix elements and dark matter profiles.  
The lightest Higgs mass is fixed to be 125.1 GeV,
and the squark, slepton, gluino, and $H,A, H^\pm$ Higgs boson masses are set to 10 TeV.
The lighter (red) parts of the model lines correspond to the nominal ATLAS exclusion
region \cite{ATLAS:2021yqv} for pair-produced winos decaying to higgsinos, although the decay branching ratios and kinematics differ.
}
\end{figure}

To this end, in Figure \ref{fig:DeltaM0} I show
the ATLAS and CMS expected and observed limits in the plane
of $\Delta M_0$ and $M_{\tilde N_2}$. The region of excesses has $\Delta M_0$ 
larger than about 10 GeV. As we have seen in Figures \ref{fig:deltaMp} and \ref{fig:maxDeltam}, accommodating sizeable mass splittings consistently
with LZ 2022 limits, in MSSM models with sfermions, the gluino, and heavy Higgs bosons decoupled, requires negative $\mu$ and small $\tan\beta$. Accordingly, I display
some selected model lines with negative $\mu$ and $\tan\beta = 2$ (left panel) and $\tan\beta=1.5$ (right panel), focusing on the region with $M_{\tilde N_2}$ less than 270 GeV, with curves along constant values of $M_1/|\mu|$ as labeled. The solid portions of these lines obey the nominal LZ 2022 constraints, and the dashed portions relax the spin-dependent LSP-neutron cross-section by a factor of 2,
in recognition of the significant systematic uncertainties in nuclear matrix elements
and dark matter profiles. In these plots I assumed $M_2 = 1.8 M_1$, as motivated by
gaugino mass unification at high scales. 
 
%%%%%%%%%%%%%%%%%%%%%%%%%%%%%%%%%%%%%%%%%%%%%%%%%%%%%%%%%%%%%%%%%%%%%%%%%%%%%%%%%
%%%%%%%%%%%%%%%%%%%%%%%%%%%%%%%%%%%%%%%%%%%%%%%%%%%%%%%%%%%%%%%%%%%%%%%%%%%%%%%%%

It is apparent from the plots in Figure \ref{fig:DeltaM0} that these models may accommodate
the LHC excess region within the LZ 2022 bounds, but only for $\Delta M_0$ in the lower part of the excess region, with $\Delta M_0$ at most about 20 GeV. Some caution is in order here, because the published experimental limits shown here assumed $\Delta M_+/\Delta M_0 = 0.5$, while I find that  the model predictions for $M_2 = 1.8 M_1$ instead give this ratio in the range from about 0.65 to 0.75. (More general ratios are possible for different assumptions about $M_2/M_1$.) The production cross-section for $\tilde C_1 \tilde N_2$ will therefore be slightly lower in the actual models than in the experimental exclusion assumptions, while the chargino decay products will have higher energy. Therefore, the true exclusion region that would follow from a dedicated analysis of the experimental data (certainly beyond the scope of this paper) might well differ from that shown. The production cross-sections
for higgsinos are very roughly in the right range (a substantial fraction of a pb), but it seems premature to attempt a more definitive statement about fitting rates in light of the quite tentative nature of the excesses. 

\begin{figure}[!t]
\mbox{
\includegraphics[width=0.52\linewidth]{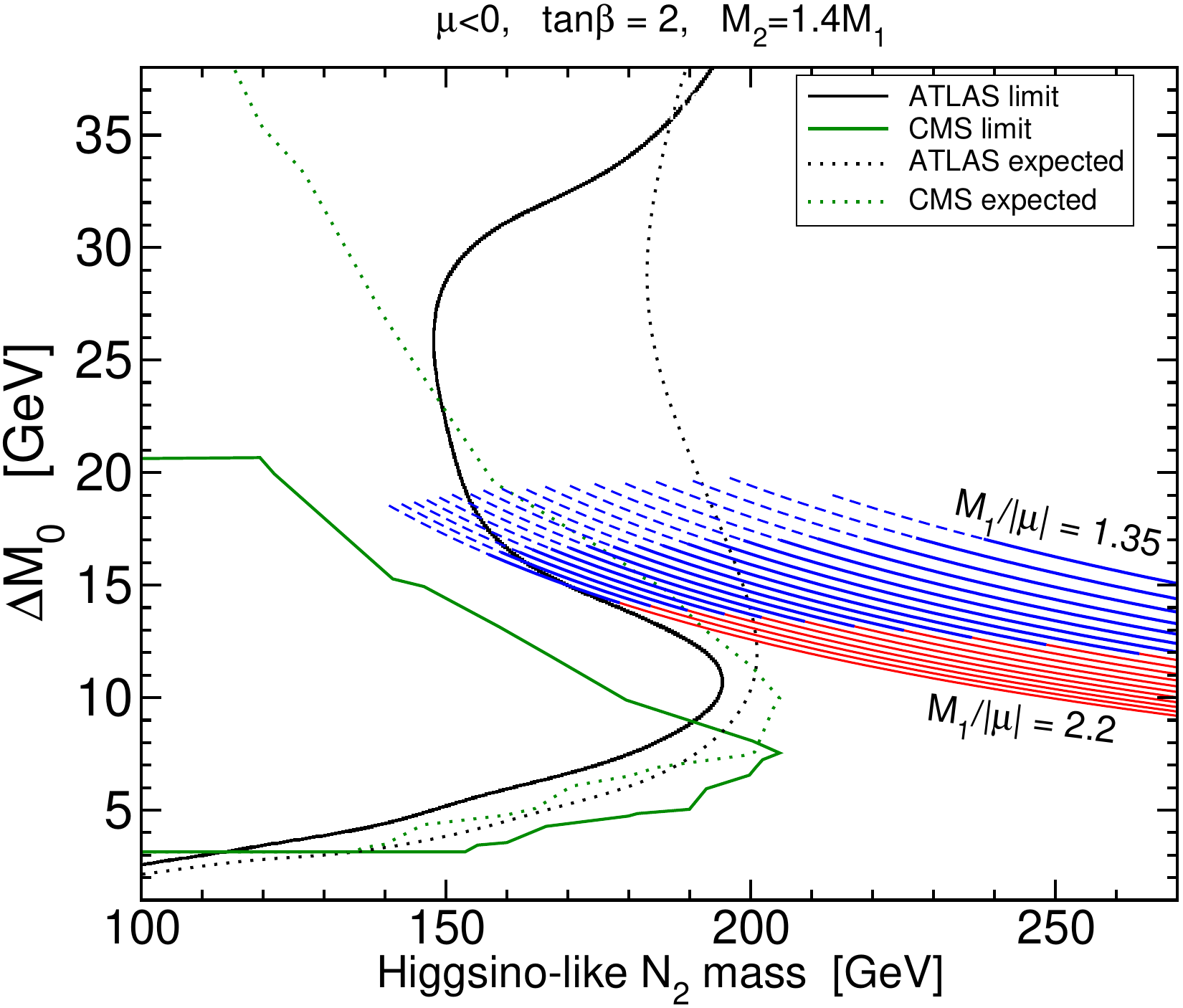}
\includegraphics[width=0.52\linewidth]{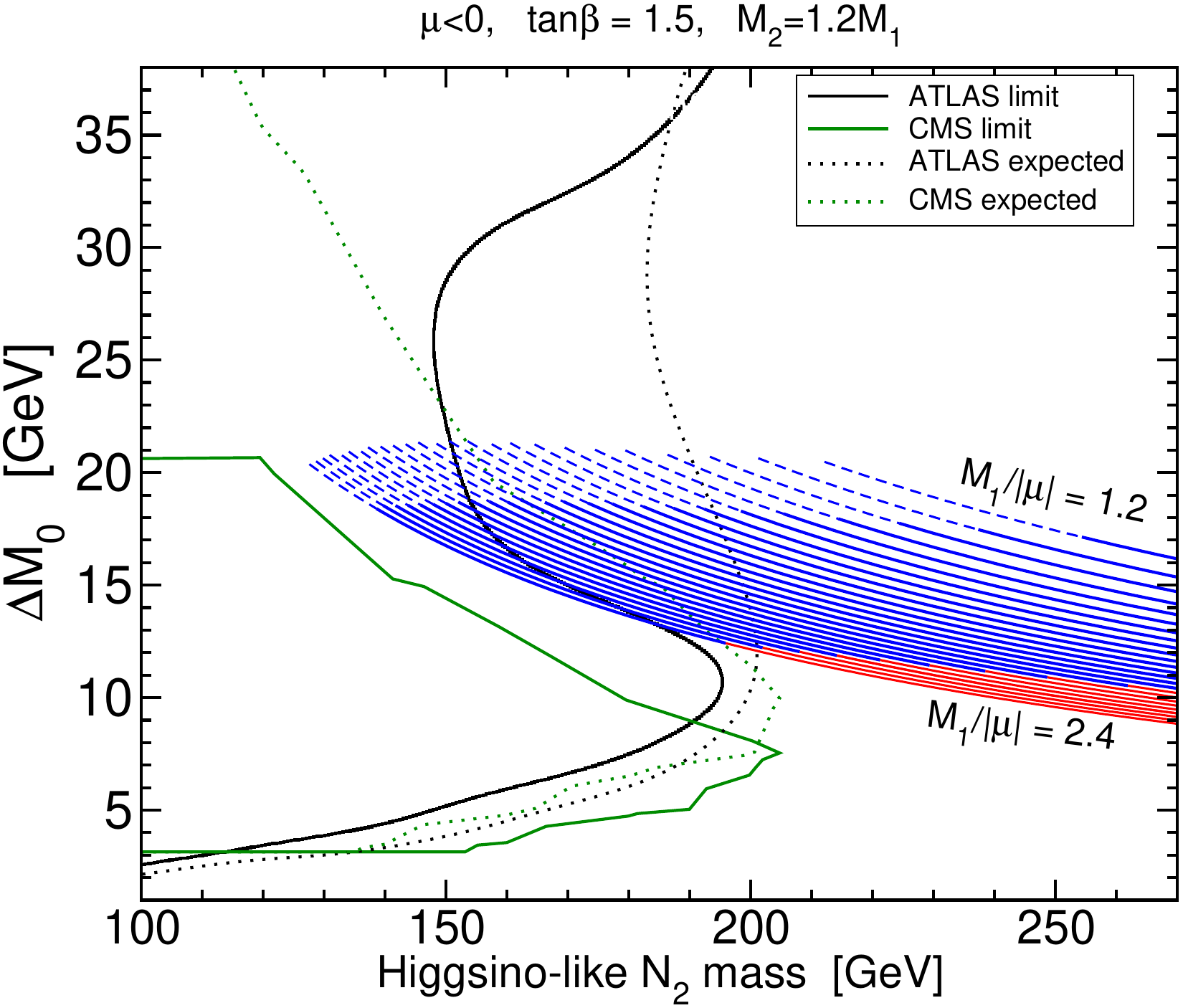}
}
\caption{\label{fig:DeltaM0moar}
The higgsino-like mass splitting $\Delta M_0 = M_{\tilde N_2} - M_{\tilde N_1}$,
as a function of $M_{\tilde N_2}$, for models with $\mu<0$ and small $\tan\beta$.
Also shown are the ATLAS \cite{ATLAS:2021moa} and CMS \cite{CMS:2021edw} higgsino soft lepton search limits (solid curves) and expected limits (dotted curves) assuming $M_{\tilde C_1} = 0.5 (M_{\tilde N_1} + M_{\tilde N_2})$.
The model lines in the left panel have $\tan\beta=2$ and $M_2 = 1.4 M_1$ with $M_1/|\mu| = 1.35$ to 2.2 in increments of 0.05, while
the right panel has $\tan\beta=1.5$ and $M_2 = 1.2 M_1$ with $M_1/|\mu| = 1.2$ to 2.4 in increments of 0.05.
The solid portions of the model lines are allowed by the nominal LZ 2022 \cite{LZ:2022lsv} direct detection searches, while the dashed portions relax the spin-dependent LSP-neutron cross-section limits by a factor of 2, in recognition of the significant uncertainties in neutron matrix elements and dark matter profiles. 
The lightest Higgs mass is fixed to be 125.1 GeV,
and the squark, slepton, gluino, and $H,A, H^\pm$ Higgs boson masses are set to 10 TeV.
The lighter (red) parts of the model lines correspond to the nominal ATLAS exclusion region  \cite{ATLAS:2021yqv} for pair-produced winos decaying to higgsinos, although the decay branching ratios and kinematics differ. }
\end{figure}

An interesting feature of these models is that the wino-like states (which are $\tilde N_4$ and $\tilde C_2$, assuming the mass ordering $|\mu| < M_1 < M_2$ as here) may also be within the kinematic reach of the LHC, and therefore can also be a promising target for searches. The winos decay have only suppressed couplings to the lighter bino, and so decay mostly to the higgsino-like states, with 
\beq
\tilde N_4 &\rightarrow& W^\pm \tilde C_1^\pm,\qquad {\rm BR} = 0.51
\\
\tilde N_4 &\rightarrow& Z \tilde N_{1,2},\qquad {\rm BR} = 0.27
\\
\tilde N_4 &\rightarrow& h \tilde N_{1,2},\qquad {\rm BR} = 0.19
\eeq
and
\beq
\tilde C_2 &\rightarrow& h \tilde C_1,\qquad {\rm BR} = 0.48
\\
\tilde C_2 &\rightarrow& W \tilde N_{1,2},\qquad {\rm BR} = 0.33
\\
\tilde C_2 &\rightarrow& Z \tilde C_1,\qquad {\rm BR} = 0.17
\eeq
where the numerical values shown are for the case $\tan\beta =2$, with $\mu = -180$ GeV, $M_1 = 270$ GeV, and $M_2 = 480$ GeV at the renormalization scale $Q=10$ TeV, but significant variations occur for other parameters. LHC search bounds for pair-produced wino-like states decaying to higgsino-like states can be found in ref.~\cite{ATLAS:2021yqv}, but neglecting the effects of bino mixing 
and using $\tan\beta=10$ and $\mu>0$. Although this is not directly comparable because of the different branching ratios and kinematics, and is an indirect constraint since it involves the winos rather than the higgsinos whose masses are plotted in Figure \ref{fig:DeltaM0}, I have used a lighter red line to indicate the portions of the model lines that are nominally in the exclusion region of ref.~\cite{ATLAS:2021yqv}. (It is claimed in ref.~\cite{ATLAS:2021yqv}
that the limits are not overly sensitive to variations in $\tan\beta$ and sign$(\mu)$, despite significant variations in the wino branching ratios.)
Bounds on winos decaying to higgsinos are more easily evaded when $M_2$ is not too large compared to $|\mu|$; this gives a larger mass splitting among the higgsinos, but crucially a more compressed wino-higgsino mass splitting.
Accommodation of the soft lepton excess is therefore found to be easier for smaller $\tan\beta$ and smaller gaugino mass ratio $M_2/M_1$. This is illustrated in Figure \ref{fig:DeltaM0moar}, in which I show selected model lines for $\tan\beta=2$ with $M_2/M_1=1.4$, and for $\tan\beta=1.5$ with $M_2/M_1 = 1.2$. Even smaller ratios $M_2/M_1$ may further enlarge the relevant parameter space. Avoiding possible constraints from wino pair production can also be done by simply choosing $M_2$ larger than 1 TeV, but I find that although this may still
accommodate the soft lepton excess, it is more difficult
because it decreases the mass splittings among the higgsinos. Therefore, continuing searches for winos decaying to higgsinos should be of particular interest. This should include searches for same-sign-$W$ pairs without a large amount of hadronic activity or $b$-tags, coming from $pp \rightarrow \tilde N_4 \tilde C_2$ followed by $\tilde C_2 \rightarrow W \tilde N_{1,2}$ and $\tilde N_4 \rightarrow W \tilde C_1$, as suggested in ref.~\cite{Baer:2013yha}.

%%%%%%%%%%%%%%%%%%%%%%%%%%%%%%%%%%%%%%%%%%%%%%%%%%%%%%%%%%%%%%%%%%%%%%%%%%%%%%%%%
%%%%%%%%%%%%%%%%%%%%%%%%%%%%%%%%%%%%%%%%%%%%%%%%%%%%%%%%%%%%%%%%%%%%%%%%%%%%%%%%%

It should also be noted that it is not mandatory to impose the LZ 2022 dark matter constraints at all. For example, a light higgsino that is apparently the LSP could actually decay on a length scale far larger than the physical size of the collider detectors, for example to a gravitino/goldstino, or to a lighter axino, or to Standard Model states via small $R$-parity violating interactions. In that case,
with no dark matter constraints to implement, it is much easier to accommodate
the mass splittings that correspond to the soft lepton excess region. This would correspond, for example, to the dashed portions of the lines in Figure \ref{fig:deltaMp},
with no obvious preference for particular values of $\tan\beta$ or sign$(\mu)$. I know of
no reason to discount this possibility.

At the other extreme, one could consider the possibility that all of the
dark matter is the higgsino-like LSP with $\Omega_{\rm LSP} h^2 = 0.12$.
For masses much less than 1.1 TeV, this would require some non-thermal source, such as out-of-equilibrium decays of some heavier particle(s) after freezeout but before nucleosynthesis \cite{Moroi:1999zb,Gelmini:2006pw}. This is an attractive possibility on general grounds, but with $\xi=1$, and therefore no density suppression of the spin-independent or spin-dependent LSP nucleon cross-sections, the LZ 2022 direct detection constraints on higgsino purity become much more severe, and it seems very unlikely to be able to have higgsino mass splittings large enough to accommodate the LHC soft lepton excess.

The discussion of the soft lepton excess and the potentially overlapping monojet excess in ref.~\cite{Agin:2023yoq} takes the option of not imposing requirements based on dark matter. A very recent paper \cite{Chakraborti:2024pdn} imposes dark matter constraints in a similar way to that done here, but with somewhat different assumptions about parameters, and so comes to a different conclusion regarding the possibility of accommodating it with light higgsinos. In that paper, $\mu$ and $M_2$ are assumed to have the same sign,
and the consistency with the $\sigma_{\rm SI}$ limit from LZ 2022 relies on an amplitude cancellation \cite{Baum:2023inl}
between neutral Higgs $h$ and $H$ contributions, which, as pointed out in those papers, entails other debilitating constraints from $H/A \rightarrow \tau^+\tau^-$ and the observed Standard-Model-like nature of $h$ production and decay. In contrast, in this paper I have assumed that the heavy Higgs bosons $H, A, H^\pm$ are all safely and thoroughly decoupled,\footnote{Of course, relaxing this assumption could further enlarge the parameter space consistent with the LZ 2022 limits and relevant to the soft lepton excesses.} and the required suppression of $\sigma_{\rm SI}$ comes about from having low $\tan\beta$ and negative $\mu$, which can be seen as due to the $1 +\sigma s_{2 \beta}$ factor in $y_h$ in eq.~(\ref{eq:yh}). In the region of parameter space critical to the soft lepton excess, the limiting factor on the mass splittings therefore usually arises from the less constrained, but less suppressed, $\sigma_{\rm SD}^n$.

\baselineskip=15.25pt

\section{Outlook\label{sec:outlook}}
\setcounter{equation}{0}
\setcounter{figure}{0}
\setcounter{table}{0} 
\setcounter{footnote}{1}

In this paper, I have examined the implications of dark matter direct detection bounds
from LZ 2022 on the purity of light higgsinos, assuming that the LSP is stable and makes up at least a fraction of the dark matter, with a density set by thermal freezeout. Motivated by the good agreement of the Standard Model with flavor- and CP-violating observables, I concentrated on the case that the scalars of the MSSM are heavy enough to be effectively decoupled. The resulting purity constraints give lower bounds on the gaugino masses, while allowing for a significant reach in future direct detection experiments before the neutrino fog is reached. Depending on the parameters, this reach corresponds to wino and bino masses well above 10 TeV, with the strongest present bounds and greatest future reach occurring for small $\tan\beta$ and positive $\mu$, and the weakest present bounds and future reach also occurring for small $\tan\beta$ but negative $\mu$. I used default settings in {\tt micrOMEGAs 6.0} for nuclear matrix elements, and it should be recognized that there is a significant uncertainty associated with these and with the local dark matter profile.

The same purity constraints give rise to upper bounds on the mass splittings among the higgsinos. The remaining parameter spaces are a particularly difficult challenge for the LHC. In particular, I considered the compatibility of scalar-decoupled light-higgsino models with the excess that currently exists in LHC soft lepton searches for charginos and neutralinos. In the case that the LSP is stable with a thermal relic abundance, this points to a region of parameter space with negative $\mu$ and small $\tan\beta$ (of order 3 or less). If, instead, the LSP is unstable
or has a relic abundance smaller than predicted by thermal freezeout, then accommodation
of the soft lepton excess is much easier due to the lack of a dark matter constraint, and may not prefer any particular $\tan\beta$ range or sign$(\mu)$. Of course, there can also be other completely different scenarios to explain the excess, as in \cite{Agin:2023yoq,Chakraborti:2024pdn}. In any case, more data will settle the issue.

Whether or not it makes up a portion of the dark matter, collider searches for light higgsinos should be an important component of the future LHC experimental program, especially given the meager constraints that exist at present, and subjective but interesting theoretical motivations based on naturalness (see e.g.~refs.~\cite{Baer:2012cf,Bae:2015jea,Baer:2018rhs} and references therein). Simplified models do not capture the physics variety of these models, or supersymmetric theories more generally. Detection efficiencies can be particularly sensitive to small and moderate mass splittings of realistic models of neutralinos and charginos. One possible remedy for this problem is to simply increase the variety of simplified models for which the experimental collaborations present results, so that any given realistic model is closer to some simplified model. An even more ambitious approach might be to change the paradigm by which experimental search results are presented, so that a publicly available software tool, validated and provided directly by the collaborations based on their knowledge of experimental realities and analysis methods, could provide (or decline to provide, if parameters are not in appropriate ranges) $p$-values for any input parameter file. If feasible, the case of charginos and neutralinos with decoupled scalars might be an ideal testing ground for such an approach to dissemination of search results. 
 
Acknowledgments: I thank James Wells for helpful conversations. This work is supported in part by the National Science Foundation grant with award number 2310533.

%%%%%%%%%%%%%%%%%%%%%%%%%%%%%%%%%%%%%%%%%%%%%%%%%%%%%%%%%%%%%%%%%%%%


\begin{thebibliography}{90}
\baselineskip=11.9pt

\bibitem{Martin:1997ns}
S.~P.~Martin,
``A Supersymmetry primer,''
%Adv. Ser. Direct. High Energy Phys. \textbf{18}, 1-98 (1998)
%doi:10.1142/9789812839657\_0001
[arXiv:hep-ph/9709356 [hep-ph]].

\bibitem{Dreiner:2023yus}
H.~K.~Dreiner, H.~E.~Haber and S.~P.~Martin,
``From Spinors to Supersymmetry,''
Cambridge University Press, 2023,
ISBN 978-1-139-04974-0
%doi:10.1017/9781139049740

\bibitem{Wells:2003tf}
J.~D.~Wells,
``Implications of supersymmetry breaking with a little hierarchy between gauginos and scalars,''
[arXiv:hep-ph/0306127].

\bibitem{Arkani-Hamed:2004zhs}
N.~Arkani-Hamed, S.~Dimopoulos, G.~F.~Giudice and A.~Romanino,
``Aspects of split supersymmetry,''
Nucl. Phys. B \textbf{709}, 3-46 (2005)
%doi:10.1016/j.nuclphysb.2004.12.026
[arXiv:hep-ph/0409232].

\bibitem{Wells:2004di}
J.~D.~Wells,
``PeV-scale supersymmetry,''
Phys. Rev. D \textbf{71}, 015013 (2005)
%doi:10.1103/PhysRevD.71.015013
[arXiv:hep-ph/0411041].

\bibitem{Arvanitaki:2012ps}
A.~Arvanitaki, N.~Craig, S.~Dimopoulos and G.~Villadoro,
``Mini-Split,''
JHEP \textbf{02}, 126 (2013)
%doi:10.1007/JHEP02(2013)126
[arXiv:1210.0555 [hep-ph]].

\bibitem{Roszkowski:2017nbc}
L.~Roszkowski, E.~M.~Sessolo and S.~Trojanowski,
``WIMP dark matter candidates and searches\textemdash{}current status and future prospects,''
Rept. Prog. Phys. \textbf{81}, no.6, 066201 (2018)
%doi:10.1088/1361-6633/aab913
[arXiv:1707.06277 [hep-ph]].

\bibitem{Feng:2022rxt}
J.~L.~Feng,
``The WIMP paradigm: Theme and variations,''
SciPost Phys. Lect. Notes \textbf{71}, 1 (2023)
%doi:10.21468/SciPostPhysLectNotes.71
[arXiv:2212.02479 [hep-ph]].

\bibitem{Safdi:2022xkm}
B.~R.~Safdi,
``TASI Lectures on the Particle Physics and Astrophysics of Dark Matter,''
PoS \textbf{TASI2022}, 009 (2024)
%doi:10.22323/1.439.0009
[arXiv:2303.02169 [hep-ph]].

\bibitem{Planck:2018vyg}
N.~Aghanim \textit{et al.} [Planck],
``Planck 2018 results. VI. Cosmological parameters,''
Astron. Astrophys. \textbf{641}, A6 (2020)
[erratum: Astron. Astrophys. \textbf{652}, C4 (2021)]
%doi:10.1051/0004-6361/201833910
[arXiv:1807.06209 [astro-ph.CO]].

\bibitem{Belanger:2001fz}
G.~Belanger, F.~Boudjema, A.~Pukhov and A.~Semenov,
``MicrOMEGAs: A Program for calculating the relic density in the MSSM,''
Comput. Phys. Commun. \textbf{149}, 103-120 (2002)
%doi:10.1016/S0010-4655(02)00596-9
[arXiv:hep-ph/0112278].

\bibitem{Belanger:2004yn}
G.~Belanger, F.~Boudjema, A.~Pukhov and A.~Semenov,
``micrOMEGAs: Version 1.3,''
Comput. Phys. Commun. \textbf{174}, 577-604 (2006)
%doi:10.1016/j.cpc.2005.12.005
[arXiv:hep-ph/0405253].

\bibitem{Belanger:2020gnr}
G.~Belanger, A.~Mjallal and A.~Pukhov,
``Recasting direct detection limits within micrOMEGAs and implication for non-standard Dark Matter scenarios,''
Eur. Phys. J. C \textbf{81}, no.3, 239 (2021)
%doi:10.1140/epjc/s10052-021-09012-z
[arXiv:2003.08621 [hep-ph]].

\bibitem{Alguero:2023zol}
G.~Alguero, G.~Belanger, F.~Boudjema, S.~Chakraborti, A.~Goudelis, S.~Kraml, A.~Mjallal and A.~Pukhov,
``micrOMEGAs 6.0: N-component dark matter,''
Comput. Phys. Commun. \textbf{299}, 109133 (2024)
%doi:10.1016/j.cpc.2024.109133
[arXiv:2312.14894 [hep-ph]].

\bibitem{Bae:2013bva}
K.~J.~Bae, H.~Baer and E.~J.~Chun,
``Mainly axion cold dark matter from natural supersymmetry,''
Phys. Rev. D \textbf{89}, no.3, 031701 (2014)
%doi:10.1103/PhysRevD.89.031701
[arXiv:1309.0519 [hep-ph]].

\bibitem{Bae:2014yta}
K.~J.~Bae, H.~Baer and H.~Serce,
``Natural little hierarchy for SUSY from radiative breaking of the Peccei-Quinn symmetry,''
Phys. Rev. D \textbf{91}, no.1, 015003 (2015)
%doi:10.1103/PhysRevD.91.015003
[arXiv:1410.7500 [hep-ph]].

\bibitem{Bae:2017hlp}
K.~J.~Bae, H.~Baer and H.~Serce,
``Prospects for axion detection in natural SUSY with mixed axion-higgsino dark matter: back to invisible?,''
JCAP \textbf{06}, 024 (2017)
%doi:10.1088/1475-7516/2017/06/024
[arXiv:1705.01134 [hep-ph]].

\bibitem{Baer:2019uom}
H.~Baer, V.~Barger, D.~Sengupta, H.~Serce, K.~Sinha and R.~W.~Deal,
``Is the magnitude of the Peccei\textendash{}Quinn scale set by the landscape?,''
Eur. Phys. J. C \textbf{79}, no.11, 897 (2019)
%doi:10.1140/epjc/s10052-019-7408-x
[arXiv:1905.00443 [hep-ph]].


%%%%%%%%%%%%%%%%%%%%%%%%%%%%%%%%%%%%%

\bibitem{Drees:1996pk}
M.~Drees, M.~M.~Nojiri, D.~P.~Roy and Y.~Yamada,
``Light Higgsino dark matter,''
Phys. Rev. D \textbf{56}, 276-290 (1997)
[erratum: Phys. Rev. D \textbf{64}, 039901 (2001)]
%doi:10.1103/PhysRevD.64.039901
[arXiv:hep-ph/9701219].

\bibitem{Thomas:1998wy}
S.~D.~Thomas and J.~D.~Wells,
``Phenomenology of Massive Vectorlike Doublet Leptons,''
Phys. Rev. Lett. \textbf{81}, 34-37 (1998)
%doi:10.1103/PhysRevLett.81.34
[arXiv:hep-ph/9804359].

\bibitem{Giudice:2004tc}
G.~F.~Giudice and A.~Romanino,
``Split supersymmetry,''
Nucl. Phys. B \textbf{699}, 65-89 (2004)
[erratum: Nucl. Phys. B \textbf{706}, 487-487 (2005)]
%doi:10.1016/j.nuclphysb.2004.08.001
[arXiv:hep-ph/0406088].

\bibitem{Profumo:2004at}
S.~Profumo and C.~E.~Yaguna,
``A Statistical analysis of supersymmetric dark matter in the MSSM after WMAP,''
Phys. Rev. D \textbf{70}, 095004 (2004)
%doi:10.1103/PhysRevD.70.095004
[arXiv:hep-ph/0407036].

\bibitem{Hisano:2004ds}
J.~Hisano, S.~Matsumoto, M.~M.~Nojiri and O.~Saito,
``Non-perturbative effect on dark matter annihilation and gamma ray signature from galactic center,''
Phys. Rev. D \textbf{71}, 063528 (2005)
%doi:10.1103/PhysRevD.71.063528
[arXiv:hep-ph/0412403].

\bibitem{Baer:2011ec}
H.~Baer, V.~Barger and P.~Huang,
``Hidden SUSY at the LHC: the light higgsino-world scenario and the role of a lepton collider,''
JHEP \textbf{11}, 031 (2011)
%doi:10.1007/JHEP11(2011)031
[arXiv:1107.5581 [hep-ph]].

\bibitem{Baer:2012cf}
H.~Baer, V.~Barger, P.~Huang, D.~Mickelson, A.~Mustafayev and X.~Tata,
``Radiative natural supersymmetry: Reconciling electroweak 
fine-tuning and the Higgs boson mass,''
Phys. Rev. D \textbf{87}, no.11, 115028 (2013)
%doi:10.1103/PhysRevD.87.115028
[arXiv:1212.2655 [hep-ph]].

\bibitem{Baer:2014cua}
H.~Baer, A.~Mustafayev and X.~Tata,
``Monojets and mono-photons from light higgsino pair production at LHC14,''
Phys. Rev. D \textbf{89}, no.5, 055007 (2014)
%doi:10.1103/PhysRevD.89.055007
[arXiv:1401.1162 [hep-ph]].

\bibitem{Nagata:2014wma}
N.~Nagata and S.~Shirai,
``Higgsino Dark Matter in High-Scale Supersymmetry,''
JHEP \textbf{01}, 029 (2015)
%doi:10.1007/JHEP01(2015)029
[arXiv:1410.4549 [hep-ph]].

\bibitem{Evans:2014pxa}
J.~L.~Evans, M.~Ibe, K.~A.~Olive and T.~T.~Yanagida,
``Light Higgsinos in Pure Gravity Mediation,''
Phys. Rev. D \textbf{91}, 055008 (2015)
%doi:10.1103/PhysRevD.91.055008
[arXiv:1412.3403 [hep-ph]].

\bibitem{Bae:2015jea}
K.~J.~Bae, H.~Baer, V.~Barger, M.~R.~Savoy and H.~Serce,
``Supersymmetry with radiatively-driven naturalness: implications for WIMP and axion searches,''
Symmetry \textbf{7}, no.2, 788-814 (2015)
%doi:10.3390/sym7020788
[arXiv:1503.04137 [hep-ph]].

\bibitem{Fukuda:2017jmk}
H.~Fukuda, N.~Nagata, H.~Otono and S.~Shirai,
``Higgsino Dark Matter or Not: Role of Disappearing Track Searches at the LHC and Future Colliders,''
Phys. Lett. B \textbf{781}, 306-311 (2018)
%doi:10.1016/j.physletb.2018.03.088
[arXiv:1703.09675 [hep-ph]].

\bibitem{Kowalska:2018toh}
K.~Kowalska and E.~M.~Sessolo,
``The discreet charm of higgsino dark matter - a pocket review,''
Adv. High Energy Phys. \textbf{2018}, 6828560 (2018)
%doi:10.1155/2018/6828560
[arXiv:1802.04097 [hep-ph]].

\bibitem{Baer:2018rhs}
H.~Baer, V.~Barger, D.~Sengupta and X.~Tata,
``Is natural higgsino-only dark matter excluded?,''
Eur. Phys. J. C \textbf{78}, no.10, 838 (2018)
%doi:10.1140/epjc/s10052-018-6306-y
[arXiv:1803.11210 [hep-ph]].

\bibitem{Fukuda:2019kbp}
H.~Fukuda, N.~Nagata, H.~Oide, H.~Otono and S.~Shirai,
``Cornering Higgsinos Using Soft Displaced Tracks,''
Phys. Rev. Lett. \textbf{124}, no.10, 101801 (2020)
%doi:10.1103/PhysRevLett.124.101801
[arXiv:1910.08065 [hep-ph]].

\bibitem{Baer:2020sgm}
H.~Baer, V.~Barger, S.~Salam, D.~Sengupta and X.~Tata,
``The LHC higgsino discovery plane for present and future SUSY searches,''
Phys. Lett. B \textbf{810}, 135777 (2020)
%doi:10.1016/j.physletb.2020.135777
[arXiv:2007.09252 [hep-ph]].

\bibitem{Rinchiuso:2020skh}
L.~Rinchiuso, O.~Macias, E.~Moulin, N.~L.~Rodd and T.~R.~Slatyer,
``Prospects for detecting heavy WIMP dark matter with the Cherenkov Telescope Array: The Wino and Higgsino,''
Phys. Rev. D \textbf{103}, no.2, 023011 (2021)
%doi:10.1103/PhysRevD.103.023011
[arXiv:2008.00692 [astro-ph.HE]].

\bibitem{Delgado:2020url}
A.~Delgado and M.~Quir\'os,
``Higgsino Dark Matter in the MSSM,''
Phys. Rev. D \textbf{103}, no.1, 015024 (2021)
%doi:10.1103/PhysRevD.103.015024
[arXiv:2008.00954 [hep-ph]].

\bibitem{Co:2021ion}
R.~T.~Co, B.~Sheff and J.~D.~Wells,
``Race to find split Higgsino dark matter,''
Phys. Rev. D \textbf{105}, no.3, 035012 (2022)
%doi:10.1103/PhysRevD.105.035012
[arXiv:2105.12142 [hep-ph]].

\bibitem{Baer:2021srt}
H.~Baer, V.~Barger, D.~Sengupta and X.~Tata,
``New angular and other cuts to improve the Higgsino signal at the LHC,''
Phys. Rev. D \textbf{105}, no.9, 095017 (2022)
%doi:10.1103/PhysRevD.105.095017
[arXiv:2109.14030 [hep-ph]].

\bibitem{Carpenter:2021jbd}
L.~M.~Carpenter, H.~Gilmer and J.~Kawamura,
``Exploring nearly degenerate higgsinos using mono-Z/W signal,''
Phys. Lett. B \textbf{831}, 137191 (2022)
%doi:10.1016/j.physletb.2022.137191
[arXiv:2110.04185 [hep-ph]].

\bibitem{Evans:2022gom}
J.~L.~Evans and K.~A.~Olive,
``Higgsino dark matter in pure gravity mediated supersymmetry,''
Phys. Rev. D \textbf{106}, no.5, 055026 (2022)
%doi:10.1103/PhysRevD.106.055026
[arXiv:2202.07830 [hep-ph]].

\bibitem{Baer:2022qrw}
H.~Baer, V.~Barger, D.~Sengupta and X.~Tata,
``Angular cuts to reduce the $\tau \bar \tau j$ background to the higgsino signal at the LHC,''
[arXiv:2203.03700 [hep-ph]].

\bibitem{Dessert:2022evk}
C.~Dessert, J.~W.~Foster, Y.~Park, B.~R.~Safdi and W.~L.~Xu,
``Higgsino Dark Matter Confronts 14~Years of Fermi \ensuremath{\gamma}-Ray Data,''
Phys. Rev. Lett. \textbf{130}, no.20, 201001 (2023)
%doi:10.1103/PhysRevLett.130.201001
[arXiv:2207.10090 [hep-ph]].

\bibitem{Carpenter:2023agq}
L.~M.~Carpenter, H.~Gilmer, J.~Kawamura and T.~Murphy,
``Taking aim at the wino-Higgsino plane with the LHC,''
Phys. Rev. D \textbf{109}, no.1, 015012 (2024)
%doi:10.1103/PhysRevD.109.015012
[arXiv:2309.07213 [hep-ph]].

\bibitem{Bisal:2023fgb}
S.~Bisal, A.~Chatterjee, D.~Das, S.A.~Pasha,
``Radiative Corrections to Aid the Direct Detection of the Higgsino-like Neutralino Dark Matter: Spin-Independent Interactions,''
[arXiv:2311.09937 [hep-ph]].

\bibitem{Ibe:2023dcu}
M.~Ibe, Y.~Nakayama and S.~Shirai,
``Precise estimate of charged Higgsino/Wino decay rate,''
JHEP \textbf{03}, 012 (2024)
%doi:10.1007/JHEP03(2024)012
[arXiv:2312.08087 [hep-ph]].

%%%%%%%%%%%%%%%%%%%%%%%%%%%%%%%%%%%%%%%%%


\bibitem{HESS:2016mib}
H.~Abdallah \textit{et al.} [H.E.S.S.],
``Search for dark matter annihilations towards the inner Galactic halo from 10 years of observations with H.E.S.S.,''
Phys. Rev. Lett. \textbf{117}, no.11, 111301 (2016)
%doi:10.1103/PhysRevLett.117.111301
[arXiv:1607.08142 [astro-ph.HE]].

\bibitem{HESS:2018cbt}
H.~Abdallah \textit{et al.} [HESS],
``Search for $\gamma$-Ray Line Signals from Dark Matter Annihilations in the Inner Galactic Halo from 10 Years of Observations with H.E.S.S.,''
Phys. Rev. Lett. \textbf{120}, no.20, 201101 (2018)
%doi:10.1103/PhysRevLett.120.201101
[arXiv:1805.05741 [astro-ph.HE]].

\bibitem{HESS:2022ygk}
H.~Abdalla \textit{et al.} [H.E.S.S.],
``Search for Dark Matter Annihilation Signals in the H.E.S.S. Inner Galaxy Survey,''
Phys. Rev. Lett. \textbf{129}, no.11, 111101 (2022)
%doi:10.1103/PhysRevLett.129.111101
[arXiv:2207.10471 [astro-ph.HE]].

\bibitem{Fermi-LAT:2015ycq}
A.~Drlica-Wagner \textit{et al.} [Fermi-LAT and DES],
``Search for Gamma-Ray Emission from DES Dwarf Spheroidal Galaxy Candidates with Fermi-LAT Data,''
Astrophys. J. Lett. \textbf{809}, no.1, L4 (2015)
%doi:10.1088/2041-8205/809/1/L4
[arXiv:1503.02632 [astro-ph.HE]].

\bibitem{Fermi-LAT:2015att}
M.~Ackermann \textit{et al.} [Fermi-LAT],
``Searching for Dark Matter Annihilation from Milky Way Dwarf Spheroidal Galaxies with Six Years of Fermi Large Area Telescope Data,''
Phys. Rev. Lett. \textbf{115}, no.23, 231301 (2015)
%doi:10.1103/PhysRevLett.115.231301
[arXiv:1503.02641 [astro-ph.HE]].

\bibitem{MAGIC:2016xys}
M.~L.~Ahnen \textit{et al.} [MAGIC and Fermi-LAT],
``Limits to Dark Matter Annihilation Cross-Section from a Combined Analysis of MAGIC and Fermi-LAT Observations of Dwarf Satellite Galaxies,''
JCAP \textbf{02}, 039 (2016)
%doi:10.1088/1475-7516/2016/02/039
[arXiv:1601.06590 [astro-ph.HE]].

\bibitem{Ellis:2022emx}
J.~Ellis, K.~A.~Olive, V.~C.~Spanos and I.~D.~Stamou,
``The CMSSM survives Planck, the LHC, LUX-ZEPLIN, Fermi-LAT, H.E.S.S. and IceCube,''
Eur. Phys. J. C \textbf{83}, no.3, 246 (2023)
%doi:10.1140/epjc/s10052-023-11405-1
[arXiv:2210.16337 [hep-ph]].

\bibitem{IceCube:2016dgk}
M.~G.~Aartsen \textit{et al.} [IceCube],
``Search for annihilating dark matter in the Sun with 3 years of IceCube data,''
Eur. Phys. J. C \textbf{77}, no.3, 146 (2017)
[erratum: Eur. Phys. J. C \textbf{79}, no.3, 214 (2019)]
%doi:10.1140/epjc/s10052-017-4689-9
[arXiv:1612.05949 [astro-ph.HE]].

\bibitem{LZ:2022lsv}
J.~Aalbers \textit{et al.} [LZ],
``First Dark Matter Search Results from the LUX-ZEPLIN (LZ) Experiment,''
Phys. Rev. Lett. \textbf{131}, no.4, 041002 (2023)
%doi:10.1103/PhysRevLett.131.041002
[arXiv:2207.03764 [hep-ex]].

%%%%%%%%%%%%%%%%%%%%%%%%%%%%%%%%%%%%%%%%%%%%%%%%%%%%%%%%%%%%%%%%%%%%%%%%%%%%%%%%%%

\bibitem{ATLAS:2021moa}
G.~Aad \textit{et al.} [ATLAS],
``Search for chargino\textendash{}neutralino pair production in final states with three leptons and missing transverse momentum in $\sqrt{s} = 13$~TeV pp collisions with the ATLAS detector,''
Eur. Phys. J. C \textbf{81}, no.12, 1118 (2021)
%doi:10.1140/epjc/s10052-021-09749-7
[arXiv:2106.01676 [hep-ex]].
%Also known as CERN-EP-2021-059, ATLAS-CONF-2020-015.

\bibitem{CMS:2023qhl}
 [CMS],
``Combined search for electroweak production of winos, binos, higgsinos, and sleptons in proton-proton collisions at $sqrt{s}=$ 13 TeV,''
CMS-PAS-SUS-21-008.

\bibitem{CMS:2021edw}
A.~Tumasyan \textit{et al.} [CMS],
``Search for supersymmetry in final states with two or three soft leptons and missing transverse momentum in proton-proton collisions at $ \sqrt{s} $ = 13 TeV,''
JHEP \textbf{04}, 091 (2022)
%doi:10.1007/JHEP04(2022)091
[arXiv:2111.06296 [hep-ex]].
%Also known as CMS-SUS-18-004, CERN-EP-2021-168.


\bibitem{Cheung:2012qy}
C.~Cheung, L.~J.~Hall, D.~Pinner and J.~T.~Ruderman,
``Prospects and Blind Spots for Neutralino Dark Matter,''
JHEP \textbf{05}, 100 (2013)
%doi:10.1007/JHEP05(2013)100
[arXiv:1211.4873 [hep-ph]].

\bibitem{Ellis:2000ds}
J.~R.~Ellis, A.~Ferstl and K.~A.~Olive,
``Reevaluation of the elastic scattering of supersymmetric dark matter,''
Phys. Lett. B \textbf{481}, 304-314 (2000)
%doi:10.1016/S0370-2693(00)00459-7
[arXiv:hep-ph/0001005 [hep-ph]].

\bibitem{Baer:2003jb}
H.~Baer, C.~Balazs, A.~Belyaev and J.~O'Farrill,
``Direct detection of dark matter in supersymmetric models,''
JCAP \textbf{09}, 007 (2003)
%doi:10.1088/1475-7516/2003/09/007
[arXiv:hep-ph/0305191 [hep-ph]].

%%%%%%%%%%%%%%%%%%%%%%%%%%%%%%%%%%%%%%%%%%%%%%%%%%%%%%%%%%%%%%%%%%%%%%%%%%%%%%%%%%

\bibitem{Moroi:1992zk}
T.~Moroi and Y.~Okada,
``Upper bound of the lightest neutral Higgs mass in extended supersymmetric Standard Models,''
Phys. Lett. B \textbf{295}, 73-78 (1992)
doi:10.1016/0370-2693(92)90091-H

\bibitem{Moroi:1991mg}
T.~Moroi and Y.~Okada,
``Radiative corrections to Higgs masses in the supersymmetric model with an extra family and antifamily,''
Mod. Phys. Lett. A \textbf{7}, 187-200 (1992)
doi:10.1142/S0217732392000124

\bibitem{Babu:2004xg}
K.S.~Babu, I.~Gogoladze and C.~Kolda,
``Perturbative unification and Higgs boson mass bounds,''
[arXiv:hep-ph/0410085].

\bibitem{Babu:2008ge}
K.S.~Babu, I.~Gogoladze, M.U.~Rehman and Q.~Shafi,
``Higgs Boson Mass, Sparticle Spectrum and Little Hierarchy Problem in Extended MSSM,''
Phys. Rev. D \textbf{78}, 055017 (2008)
%doi:10.1103/PhysRevD.78.055017
[arXiv:0807.3055 [hep-ph]].

\bibitem{Martin:2009bg}
S.~P.~Martin,
``Extra vector-like matter and the lightest Higgs scalar boson mass in low-energy supersymmetry,''
Phys. Rev. D \textbf{81}, 035004 (2010)
%doi:10.1103/PhysRevD.81.035004
[arXiv:0910.2732 [hep-ph]].

\bibitem{Graham:2009gy}
P.~W.~Graham, A.~Ismail, S.~Rajendran and P.~Saraswat,
``A Little Solution to the Little Hierarchy Problem: A Vector-like Generation,''
Phys. Rev. D \textbf{81}, 055016 (2010)
%doi:10.1103/PhysRevD.81.055016
[arXiv:0910.3020 [hep-ph]].

%%%%%%%%%%%%%%%%%%%%%%%%%%%%%%%%%%%%%%%%%%%%%%%%%%%%%%%%%%%%%%%%%%%%%%%%%%%%%%%%%%

\bibitem{CMS:2023mny}
A.~Hayrapetyan \textit{et al.} [CMS],
``Search for supersymmetry in final states with disappearing tracks in proton-proton collisions at $\sqrt{s}$ = 13 TeV,''
[arXiv:2309.16823 [hep-ex]].
%Also known as CERN-EP-2023-209, CMS-SUS-21-006-003, CMS-PAS-SUS-21-006.

\bibitem{ATLAS:2022rme}
G.~Aad \textit{et al.} [ATLAS],
``Search for long-lived charginos based on a disappearing-track signature using 136 fb$^{-1}$ of pp collisions at $\sqrt{s}$~=~13~TeV with the ATLAS detector,''
Eur. Phys. J. C \textbf{82}, no.7, 606 (2022)
%doi:10.1140/epjc/s10052-022-10489-5
[arXiv:2201.02472 [hep-ex]].
%Also known as  CERN-EP-2021-209, ATLAS-CONF-2021-015

\bibitem{Pierce:1994ew}
D.~Pierce and A.~Papadopoulos,
``The Complete radiative corrections to the gaugino and Higgsino masses in the minimal supersymmetric model,''
Nucl. Phys. B \textbf{430}, 278-294 (1994)
%doi:10.1016/0550-3213(94)00303-3
[arXiv:hep-ph/9403240].

\bibitem{Pierce:1996zz}
D.~M.~Pierce, J.~A.~Bagger, K.~T.~Matchev and R.~j.~Zhang,
``Precision corrections in the minimal supersymmetric standard model,''
Nucl. Phys. B \textbf{491}, 3-67 (1997)
%doi:10.1016/S0550-3213(96)00683-9
[arXiv:hep-ph/9606211].

\bibitem{Eberl:2001eu}
H.~Eberl, M.~Kincel, W.~Majerotto and Y.~Yamada,
``One loop corrections to the chargino and neutralino mass matrices in the on-shell scheme,''
Phys. Rev. D \textbf{64}, 115013 (2001)
%doi:10.1103/PhysRevD.64.115013
[arXiv:hep-ph/0104109].

\bibitem{Martin:2005ch}
S.~P.~Martin,
``Fermion self-energies and pole masses at two-loop order in a general renormalizable theory with massless gauge bosons,''
Phys. Rev. D \textbf{72}, 096008 (2005)
%doi:10.1103/PhysRevD.72.096008
[arXiv:hep-ph/0509115].

\bibitem{Allanach:2001kg}
B.~C.~Allanach,
``SOFTSUSY: a program for calculating supersymmetric spectra,''
Comput. Phys. Commun. \textbf{143}, 305-331 (2002)
%doi:10.1016/S0010-4655(01)00460-X
[arXiv:hep-ph/0104145].

\bibitem{Billard:2013qya}
J.~Billard, L.~Strigari and E.~Figueroa-Feliciano,
``Implication of neutrino backgrounds on the reach of next generation dark matter direct detection experiments,''
Phys. Rev. D \textbf{89}, no.2, 023524 (2014)
%doi:10.1103/PhysRevD.89.023524
[arXiv:1307.5458 [hep-ph]].

\bibitem{OHare:2021utq}
C.~A.~J.~O'Hare,
``New Definition of the Neutrino Floor for Direct Dark Matter Searches,''
Phys. Rev. Lett. \textbf{127}, no.25, 251802 (2021)
%doi:10.1103/PhysRevLett.127.251802
[arXiv:2109.03116 [hep-ph]].

\bibitem{Randall:1998uk}
L.~Randall and R.~Sundrum,
``Out of this world supersymmetry breaking,''
Nucl. Phys. B \textbf{557}, 79-118 (1999)
%doi:10.1016/S0550-3213(99)00359-4
[arXiv:hep-th/9810155].

\bibitem{Giudice:1998xp}
G.~F.~Giudice, M.~A.~Luty, H.~Murayama and R.~Rattazzi,
``Gaugino mass without singlets,''
JHEP \textbf{12}, 027 (1998)
%doi:10.1088/1126-6708/1998/12/027
[arXiv:hep-ph/9810442].

\bibitem{ATLAS:2024umc}
G.~Aad \textit{et al.} [ATLAS],
``Search for nearly mass-degenerate higgsinos using low-momentum mildly-displaced tracks in $pp$ collisions at $\sqrt{s}$ = 13 TeV with the ATLAS detector,''
[arXiv:2401.14046 [hep-ex]].

\bibitem{Agin:2023yoq}
D.~Agin, B.~Fuks, M.~D.~Goodsell and T.~Murphy,
``Monojets reveal overlapping excesses for light compressed higgsinos,''
[arXiv:2311.17149 [hep-ph]].

\bibitem{ATLAS:2021kxv}
G.~Aad \textit{et al.} [ATLAS],
``Search for new phenomena in events with an energetic jet and missing transverse momentum in $pp$ collisions at $\sqrt {s}$ =13  TeV with the ATLAS detector,''
Phys. Rev. D \textbf{103}, no.11, 112006 (2021)
%doi:10.1103/PhysRevD.103.112006
[arXiv:2102.10874 [hep-ex]].
%Also known as CERN-EP-2020-238.

\bibitem{CMS:2021far}
A.~Tumasyan \textit{et al.} [CMS],
``Search for new particles in events with energetic jets and large missing transverse momentum in proton-proton collisions at $ \sqrt{s} $ = 13 TeV,''
JHEP \textbf{11}, 153 (2021)
%doi:10.1007/JHEP11(2021)153
[arXiv:2107.13021 [hep-ex]].
%Also known as CMS-EXO-20-004, CERN-EP-2021-136.

%%%%%%%%%%%%%%
\bibitem{ATLAS:2021yqv}
G.~Aad \textit{et al.} [ATLAS],
``Search for charginos and neutralinos in final states with two boosted hadronically decaying bosons and missing transverse momentum in $pp$ collisions at $\sqrt {s}$ = 13\,\,TeV with the ATLAS detector,''
Phys. Rev. D \textbf{104}, no.11, 112010 (2021)
%doi:10.1103/PhysRevD.104.112010
[arXiv:2108.07586 [hep-ex]].

\bibitem{Baer:2013yha}
H.~Baer, V.~Barger, P.~Huang, D.~Mickelson, A.~Mustafayev, W.~Sreethawong and X.~Tata,
``Same sign diboson signature from supersymmetry models with light higgsinos at the LHC,''
Phys. Rev. Lett. \textbf{110}, no.15, 151801 (2013)
%doi:10.1103/PhysRevLett.110.151801
[arXiv:1302.5816 [hep-ph]].

\bibitem{Moroi:1999zb}
T.~Moroi and L.~Randall,
``Wino cold dark matter from anomaly mediated SUSY breaking,''
Nucl. Phys. B \textbf{570}, 455-472 (2000)
%doi:10.1016/S0550-3213(99)00748-8
[arXiv:hep-ph/9906527].

\bibitem{Gelmini:2006pw}
G.~B.~Gelmini and P.~Gondolo,
``Neutralino with the right cold dark matter abundance in (almost) any supersymmetric model,''
Phys. Rev. D \textbf{74}, 023510 (2006)
%doi:10.1103/PhysRevD.74.023510
[arXiv:hep-ph/0602230].

\bibitem{Chakraborti:2024pdn}
M.~Chakraborti, S.~Heinemeyer and I.~Saha,
``Consistent Excesses in the Search for $\tilde \chi_2^{\rm 0} \tilde \chi_1^{\rm \pm}$ : Wino/bino vs. Higgsino Dark Matter,''
[arXiv:2403.14759 [hep-ph]].

\bibitem{Baum:2023inl}
S.~Baum, M.~Carena, T.~Ou, D.~Rocha, N.~R.~Shah and C.~E.~M.~Wagner,
``Lighting up the LHC with Dark Matter,''
JHEP \textbf{11}, 037 (2023)
%doi:10.1007/JHEP11(2023)037
[arXiv:2303.01523 [hep-ph]].

\end{thebibliography}
\end{document}